\documentclass[journal=jacsat,manuscript=article]{achemso}
\usepackage[version=3]{mhchem} 

\usepackage{filecontents}

\usepackage{amsmath} 
\newcommand{\angstrom}{\textup{\AA}}

\usepackage{xcolor}
\usepackage{graphicx}
\usepackage{subfig}
\usepackage{gensymb}

\usepackage{boldline}

\usepackage{multirow}

\usepackage{xr}
\makeatletter
\newcommand*{\addFileDependency}[1]{
  \typeout{(#1)}
  \@addtofilelist{#1}
  \IfFileExists{#1}{}{\typeout{No file #1.}}
}
\makeatother

\newcommand*{\myexternaldocument}[1]{%
    \externaldocument{#1}%
    \addFileDependency{#1.tex}%
    \addFileDependency{#1.aux}%
}

\myexternaldocument{si}

\author{Bal\'{a}zs F\'{a}bi\'{a}n}
\email{balazs.fabian@biophys.mpg.de}
\affiliation[MPI-BP]{Department of Theoretical Biophysics, Max Planck Institute of Biophysics, Max-von-Laue Straße 3, 60438, Frankfurt am Main, Germany}

\author{Sebastian Thallmair}
\affiliation[FIAS]{Frankfurt Institute for Advanced Studies, Ruth-Moufang-Straße 1, 60438 Frankfurt am Main, Germany}

\author{Gerhard Hummer}
\affiliation[MPI-BP]{Department of Theoretical Biophysics, Max Planck Institute of Biophysics, Max-von-Laue Straße 3, 60438, Frankfurt am Main, Germany}

\title[] {Small ionic radii limit time step in Martini~3 molecular dynamics simulations}

\begin{document}

\begin{abstract}
    Among other improvements, the Martini~3 coarse-grained force field provides a more accurate description of the solvation of protein pockets and channels through the consistent use of various bead types and sizes. Here, we show that the representation of Na$^+$ and Cl$^-$ ions as ``tiny'' (TQ5) beads limits the accessible time step to 25~fs. By contrast, with Martini~2, time steps of 30-40~fs were possible for lipid bilayer systems without proteins. This limitation is relevant for, e.g., phase separating lipid mixtures that require long equilibration times. We derive a quantitative kinetic model of time-integration instabilities in molecular dynamics (MD) as a function of time step, ion concentration and mass, system size, and simulation time. With this model, we demonstrate that ion-water interactions are the main source of instability at physiological conditions, followed closely by ion-ion interactions. We show that increasing the ionic masses makes it possible to use time steps up to 40~fs with minimal impact on static equilibrium properties and on dynamical quantities such as lipid and ion diffusion coefficients. Increasing the size of the bead representing the ions (and thus changing their hydration) also permits longer time steps. The use of larger time steps in Martini~3 simulations results in a more efficient exploration of configuration space. The kinetic model of MD simulation crashes can be used to determine the maximum allowed time step whenever sampling efficiency is critical.
\end{abstract}

\section{Introduction}

Compared to its predecessors\cite{marrink2004coarse, marrink2007martini}, the recent Martini~3 force field\cite{souza2021martini} constitutes a significant advance in biomolecular simulations. Martini~3 offers consistently parametrized coarse-grained (CG) interaction sites of different size and a better coverage of the chemical space by a whole range of new bead types. This results in an improved representation of molecular shape, packing, and interactions in general. The diversity of the ion models was strongly increased, now featuring five bead types Q1-Q5 in three resolutions ranging from ``tiny'' (T, 2-to-1 mapping) via ``small'' (S, 3-to-1 mapping) to ``regular'' (R, 4-to-1 mapping) beads. The parametrization of the beads features a Martini Hofmeister series from hard, more inorganic ions (Q5) to soft, more organic ions (Q1). In addition, specific interactions (e.g., of cation-$\pi$ type) can be incorporated using labels which were previously only available in the polarizable version. \cite{khan2020capturing} The new features of Martini~3 resulted in a change of the resolution of hard ions such as Na$^+$ and Cl$^-$ which are routinely added to biomolecular simulation systems to establish physiological conditions in terms of ionic strength. The size of both Na$^+$ and Cl$^-$ ions changed from R to T, which means that in Martini~3 the ions are modeled without a hydration shell. This is in contrast to Martini~2, where a hydration shell was considered to be included in the larger R-bead type used for Na$^+$ and Cl$^-$.

While Martini~2 simulations were routinely run with time steps of $\Delta t=30$~fs \cite{weiner2018presence, rosetti2017molecular, perlmutter2011interleaflet} according to the recommendation of the \textit{new-rf} parameter set \cite{de2016martini}, the current recommended time step is $\Delta t =20$~fs, which was used for all test systems during the parametrization of the force field. In several previous studies employing Martini~2 even $\Delta t=40$~fs was used \cite{marrink2007martini, xing2009asymmetric, jalili2009coarse}. However, this was deemed excessive by the developers of the model \cite{marrink2010comment}, and the use of such large time steps is generally discouraged.

Here, we show that the representation of Martini~3 ions as ``tiny'' charged beads of type TQ5 requires integration time step no larger than 25~fs. For $\Delta t > 25$~fs the tendency of the time integration to become unstable and crash the simulations increases significantly. We develop a quantitative kinetic model of the rate of crashing as a function of integration time step, ion concentration, and ionic mass. By fitting the three parameters of this model globally to the observed crash statistics of the MD runs, we identify ion-ion and ion-water interactions as the main culprits. We explore two possible solutions that permit simulations to run with larger time steps: by modifying the ionic mass and by changing the bead type of the ions. While the former keeps the desired resolution with almost no perturbations to the system, the latter is consistent with the multi-resolution approach of Martini~3. The effects and trade-offs made by these solutions are evaluated and discussed in detail. Moreover, the statistical model employed here to estimate the rate of crashing can be applied to determine an optimal time step for MD-based high-throughput campaigns. The formalism used here to determine the maximum allowed time step should prove useful whenever sampling efficiency is critical.

\section{Theory}

\subsection{Altering the mass or bead type}

In classical molecular dynamics (MD) simulations, Newton's equations of motion (usually augmented with thermostats and barostats) are integrated numerically with finite time steps. The integration time step $\Delta t$ is chosen in a trade off between accuracy, e.g., to conserve the total energy, and the computational efficiency of sampling configuration space. To maximize efficiency, the time step is commonly chosen close to the stability limit of time integration. \cite{lechner2006equilibrium} For classical non-polarizable models, this limit is determined by the fastest molecular motions, in particular the rattling of stiff covalent bonds and tight non-covalent interations, and the hard collisions between molecules on highly anharmonic potential surfaces.

For NaCl solutions and lipid bilayers, we observed that Martini~3 MD simulations required shorter time steps than in equivalent setups simulated with the Martini~2 model. As the main culprit, we identified the newly introduced ions modeled as ``tiny'' charged (TQ5) beads and their ion-ion interactions. As possible remedies, we consider (i) changing the ionic masses and (ii) changing the bead type from tiny to small (SQ5) or regular (RQ5).

The motivation behind changing the ionic mass is the invariance of Newton's equations of motion under a uniform scaling of all masses and of time by $\alpha$ and $\sqrt{\alpha}$, respectively. Whereas scaling all particle masses does not improve configuration space sampling, we expect that increasing the ionic mass will permit a longer time step. The displacements of heavier ions at each time step are smaller, which should reduce the risk of an uncontrolled clash causing an instability in the numerical time integration. If such ion displacements are limiting the time step, we expect that the maximum allowed time step for stable time integration scales with the ionic mass $m$ as
\begin{align}
    \label{eq:timestep}
    \frac{\Delta t'}{\Delta t}\approx \sqrt{\frac{m'}{m}}
\end{align}
where we ignored the motions of the other particles. Then, starting from a time step $\Delta t = 20$~fs, doubling and quadrupling the ionic mass $m$ would make it possible to use $\Delta t' = 30$ and 40~fs, respectively.

Importantly, changes in particle mass leave the equilibrium structural properties unchanged because the classical mechanical partition function separates into kinetic and configurational contributions. However, the dynamics of the system is generally modified in a nontrivial manner. Nevertheless, with a longer time step one can expect a faster sampling of the relevant configuration space for the molecules with unmodified masses.

In Martini simulations, usually only relatively slow diffusive motions are of interest, because the fast intramolecular and collision dynamics is impacted significantly by coarse-graining. Translational and rotational diffusion are dominated by hydrodynamic effects, as reflected in the Stokes--Einstein theory of 3D diffusion \cite{einstein1905molekularkinetischen} and in the Saffman--Delbr\"uck theory of 2D diffusion in lipid bilayers embedded in a 3D solvent \cite{saffman1975brownian}. Both theories predict that the diffusion coefficient of a dissolved molecular species does not depend directly on its mass, only on its size and the viscosity of the surrounding media. Therefore, we expect that ionic mass changes have minimal impact on the diffusive dynamics.

Nevertheless, changes in the mass of some of the particles can potentially change the viscosity of the aqueous solvent and the membrane. Here, we consider increasing the mass of the ions described by TQ5 beads, leaving all other masses untouched. Importantly, these particles constitute only a tiny fraction of the molecules in the aqueous solvent and do not enter within the lipid bilayers. Therefore, we expect that doubling their mass will have only a very limited effect on the viscosity of solvent and membrane, and thus in turn on the diffusive dynamics of all species, including the ions themselves. We test this assumption by monitoring the diffusivity of solvent and lipid species as a function of ionic mass. By keeping the box dimensions constant in this comparison, we do not need to correct for the large finite size effects in the computed diffusion coefficients \cite{yeh2004system,vogele2018hydrodynamics}.

Modification of the bead type of the ions is based on the notion that in Martini~3, a change in the bead size of the ions represents a change in their hydration. Bead-size changes are thus consistent with the coarse-graining philosophy of the force field (see Martini~3 Supplementary Information \cite{souza2021martini}). Moreover, the current resolution of lipid head groups such as the choline moiety of phosphatidyl-choline (PC) lipids is already low due to the employed 6-1 mapping \cite{marrink2007martini,wassenaar2015computational}, where 6 non-hydrogen atoms are grouped together into a single interaction site. Therefore, adequate modeling of protein-less bilayers currently does not require a high resolution of the ions. We thus expect that an increase in the ion bead size will not significantly perturb the membrane properties.

\subsection{Crash probability as function of time step}

In the following, we develop a kinetic model of the probability that a simulation with a given time step will crash during a given total simulation time. We assume that, rarely, the position updates in time integration place particles within a strongly repulsive region of the potential surface, such that the subsequent force evaluation leads to numerical instabilities either immediately or after a few uncontrolled further integration steps.

Let $\Delta t$ be the integration time step. With $v$ the velocity in one dimension, the corresponding displacement is then $\Delta x=v\Delta t$, ignoring the higher-order acceleration terms. We now assume that displacements $\Delta x>\Delta x_{\mathrm{crit}}$ are critical, and lead to crashes. The velocities of a particle of mass $m$ at inverse temperature $\beta$ satisfy a Maxwell-Boltzmann distribution, $p(v)dv=\exp(-\beta mv^2/2)\sqrt{\beta m/(2\pi)}dv$. The probability of a critical displacement during a simulation time step becomes
\begin{align}
    P(\Delta x>\Delta x_{\mathrm{crit}})&=\int_{\Delta x_{\mathrm{crit}}/\Delta t}
    ^\infty dv\,\exp(-\beta m v^2/2)\sqrt{\beta m/(2\pi)}\nonumber\\
    &\approx \frac{\gamma \Delta t\exp(-1/(2\gamma^2\Delta t^2))}{\sqrt{2\pi}}
\end{align}
with $1/\gamma^2=\beta m \Delta x_{\mathrm{crit}}^2$. Here we used the asymptotic expansion of the complementary error function. In a simulation system, we expect that a certain fraction $\sqrt{2\pi}f$ of the particle displacements (here, those of the ``tiny'' ions and their interaction partners) can result in such catastrophic clashes. The probability $Q=\sqrt{2\pi}fP(\Delta x>\Delta x_{\mathrm{crit}})$ of a catastrophic displacement in a given time step is approximately
\begin{equation}
    Q=f \gamma \Delta t\exp(-1/(2\gamma^2\Delta t^2))
\end{equation}
Assuming uncorrelated events and $0<Q\ll 1$, the probability that a simulation run of $n$ time steps will not crash is
\begin{equation}
    P=(1-Q)^n\approx\exp\left[-tf\gamma\exp(-1/(2\gamma^2\Delta t^2))\right]
\end{equation}
with $t=n\Delta t$ the total time. The crash times are thus predicted to be distributed exponentially, $p(t_{\text{crash}})dt_{\text{crash}}=k_{\mathrm{crash}}\exp(-k_{\mathrm{crash}}t_{\text{crash}})dt_{\text{crash}}$ with a rate that depends on the simulation time step as
\begin{equation}
    \label{eq:crash-rate}
    k_{\mathrm{crash}}(\Delta t)=f\gamma \exp(-1/(2\gamma^2\Delta t^2))=k_{\mathrm{crash}}^0 \exp(-1/(2\gamma^2\Delta t^2))
\end{equation}
The crash rate has two parameters, $f$ and $\gamma$, whose product defines $k_{\mathrm{crash}}^0=f\gamma$. In our systems, we expect that the dimensionless factor $f$ depends on the concentration $c_\mathrm{ion}$ of ions as $f=(a{c_{\mathrm{ion}}}^2+bc_{\mathrm{ion}})V$ with coefficients $a$ and $b$, and $V$ the system volume. Based on the definition, and consistent with Eq.~(\ref{eq:timestep}), we expect $\gamma$ to depend on the mass of the ions approximately as $\gamma\propto 1/\sqrt{m}$.

\section{Methods}

\subsection{Simulation of NaCl solutions}

To demonstrate the effect of changing ionic masses (or the lack thereof) on the viscosity $\mu_\mathrm{f}$ of the fluid surrounding the bilayer, we prepared water boxes with 0, 0.075, 0.15 and 0.3~M NaCl concentrations using \textit{insane.py}\cite{wassenaar2015computational} and the option \texttt{-salt}. All cubic boxes had an initial edge length of 10~nm. We simulated these NaCl solutions with the standard Martini~3 CG force field\cite{souza2021martini} and the Gromacs package\cite{abraham2015gromacs} (version 2020.1). To eliminate differences stemming from finite size effects between these systems\cite{yeh2004system}, we took the smallest box as reference and removed the appropriate number of molecules from all others. This procedure resulted in equilibrium box lengths between 8.93 and 8.97~nm for all NaCl solution systems. No corrections were applied to account for finite size effects on diffusion. At every concentration, we performed simulations with time steps of 20, 30 and 40~fs, and using the original, doubled and quadrupled ionic masses (denoted as 1$\times$m, 2$\times$m, 4$\times$m, respectively) to systematically test the effects on the diffusion coefficient, and therefore on the viscosity of the bulk medium. Following a short energy minimization using a steepest descent algorithm, every system was equilibrated during a 1~\textmu{}s run with a 20~fs time step. This was followed by 3~\textmu{}s-long production runs using 20, 30 or 40~fs time steps with writing the positions to disk every 0.5~ns. The temperature of the systems was kept at 310~K using a velocity re-scale thermostat \cite{bussi2007canonical}. The pressure was maintained at 1~bar by an isotropic Parrinello--Rahman barostat \cite{parrinello1981polymorphic} (p$_\mathrm{ref} = 1$~bar and $\tau_\mathrm{p} = 12$~ps).

To gain insight into the origin of the simulation crashes, we performed a large number of additional simulations using various combinations of concentration (0.075~M and in the range of 0.1 to 0.53~M with steps of approximately 0.05~M) and time step (every fs from 29 to 32~fs). In this set, every simulation used the settings as described above, but with a fixed number of desired steps ($n_\mathrm{steps} = 10^8$) instead of fixed total simulation time, and without trying to match the overall volumes. The majority of these simulations could not achieve the required $n_\mathrm{steps} = 10^8$, and the only observable extracted from them was the rate of crashing, as described below. To reduce the statistical uncertainty in the evaluation of crash rate, every simulation was repeated 40 times.

\begin{figure}[htb]
\begin{center}
    \includegraphics[width=10cm]{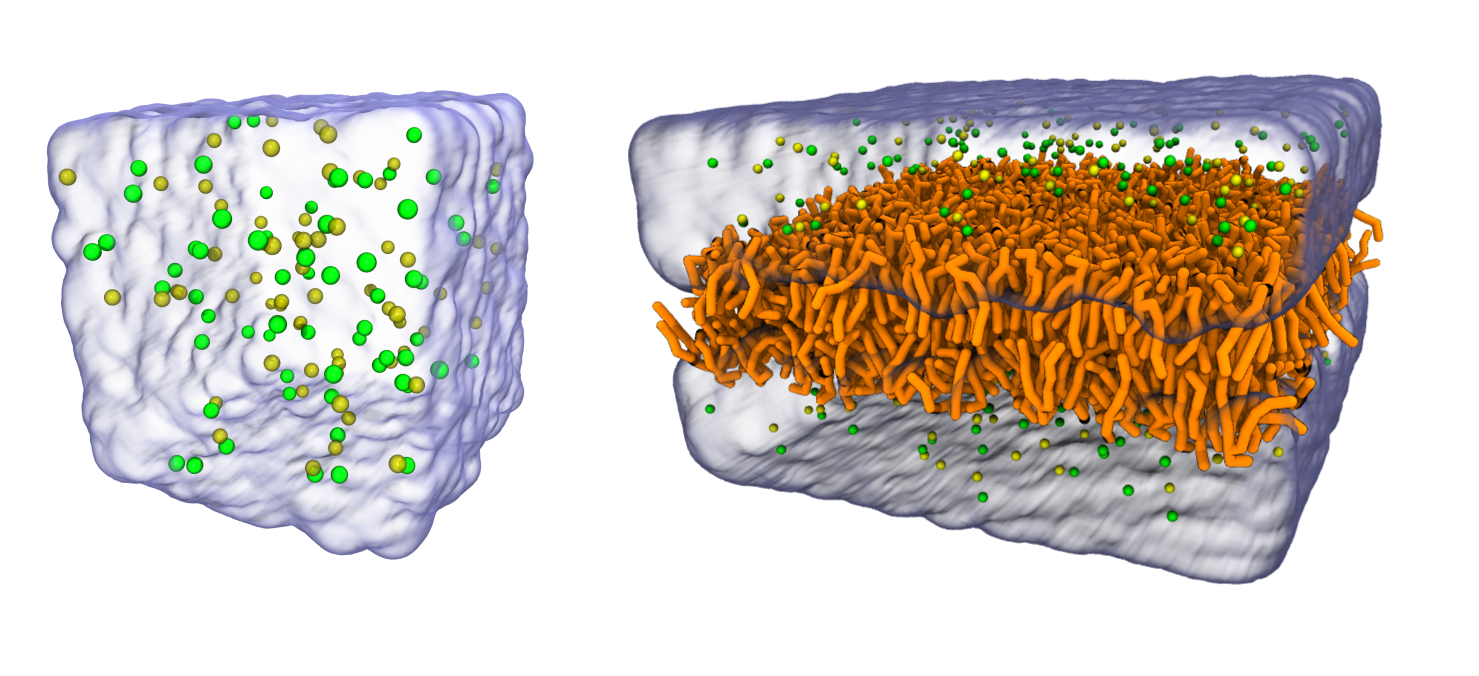}
    \caption{Snapshots of the NaCl solution system (LEFT) and the DPPC lipid bilayer system (RIGHT). Sodium and chloride ions are shown as yellow and green spheres, and DPPC lipids as orange tubes. Water is represented as transparent surface. The images were rendered using VMD \cite{HUMP96}.
    }
    \label{fig:snapshot}
\end{center}
\end{figure}

\subsection{Simulation of DPPC bilayers}

We also performed MD simulations of fully solvated DPPC bilayers containing Na$^+$ and Cl$^-$ ions with the Martini~2 and Martini~3 CG force fields. The initial systems were prepared using \textit{insane.py}\cite{wassenaar2015computational}. All bilayers were constructed in an initial box of 18$\times$18$\times$18~nm$^3$ and solvated by 14405 CG waters (pure W for Martini~3, 10\% WF antifreeze\cite{marrink2007martini} for Martini~2). The equilibrium membrane area in the simulation boxes was about 17x17~nm$^2$, larger than the $\approx$8x8~nm$^2$ used in the Martini~3 validation test. Every simulation system contained 0.15~M NaCl, as added with \textit{insane.py} using the flag \texttt{-salt 0.15}. We performed all simulations with Gromacs 2020.1, and used 20, 30 and 40~fs time steps. We simulated a Martini~2 membrane for reference, and Martini~3 membranes using TQ5 (standard Martini~3), SQ5, RQ5(=Q5) beads as monovalent ions. Finally, we tested the effect of increasing the ionic mass by a factor of 2. The doubled mass corresponds to using the interactions of TQ5 along with the mass of RQ5. The abbreviated labels of the above lipid simulations are M2, M3(-TQ5), M3-SQ5, M3-RQ5 and M3-2$\times$m, respectively. All membranes were energy minimized using a steepest descent algorithm, and further minimized by running 50~ns trajectories using 5~fs time steps. This was followed by a 5~\textmu{}s equilibration using a 20~fs time step. Then, production runs of 10~\textmu{}s were performed using 20, 30 and 40~fs time steps and writing the positions to disk every 0.5~ns. The temperature of the systems was kept at 310~K by a velocity re-scale thermostat\cite{bussi2007canonical}. The pressure of the systems was controlled in a semi-isotropic manner with a compressibility of $3\times 10^{-4}$~bar$^{-1}$. For the initial energy minimization, we employed the Berendsen barostat\cite{berendsen1984molecular} (p$_\mathrm{ref} = 1$~bar and $\tau_\mathrm{p} = 1$~ps), while the equilibration and production proceeded with the Parrinello-Rahman barostat\cite{parrinello1981polymorphic} ($p_\mathrm{ref} = 1$~bar and $\tau_\mathrm{p} = 12$~ps). Figure~\ref{fig:snapshot} shows illustrative snapshots of the simulated NaCl solutions and lipid bilayers.

\subsection{Analysis}

In our numerical tests, we ran 40 simulations for each system. 
The time of $t_i$ is the targeted run length, $t_i=\Delta t \times 10^8$, if run $i$ completed normally, or the time of the crash caused by an instability. For the exact definition of a crash, see section 1.1 of SI. From the times $t_i$, we estimate the rate of crashing (and thus the reciprocal of the mean time to a crash) using a maximum-likelihood estimator for right-censored data and Poisson statistics,
\begin{equation}
  \label{eq:crash-estimate}
  k_\mathrm{crash} = \frac{1}{t_\mathrm{crash}} = \frac{n_\mathrm{crash}}{\sum_{i=1}^{40} t_i},
\end{equation}
where $0\le n_\mathrm{crash}\le 40$ is the number of crashes observed in the 40 runs. We assessed the uncertainty of the estimator using the standard error $\sigma = k_\mathrm{crash} / \sqrt{n_\mathrm{crash}}$ which corresponds to the Cram\'{e}r--Rao bound. This estimator gives us some guidance on the molecular factors causing the crashes. For instance, if ion-water interactions were dominant, we would expect $k_\mathrm{crash}\propto c_\mathrm{ion}$, where $c_\mathrm{ion}$ is the ion concentration. By contrast, if ion-ion interactions were dominant, we would expect $k_\mathrm{crash}\propto (c_{\mathrm{ion}})^2$, as discussed in the Theory section.

From the trajectories of the bulk NaCl solutions, we determined the diffusion coefficients $D_\mathrm{3D}$ of water. We computed the diffusion coefficients with an optimal Generalized Least Squares (GLS) estimator \cite{bullerjahn2020optimal}, as implemented in the DiffusionGLS \cite{buelow-diffusiongls} package. The trajectories were unwrapped using a scheme that correctly takes into account volume fluctuations in the NPT ensemble \cite{buelow2020systematic}, as implemented in qwarp \cite{henine-qwrap}.

We evaluated the properties of the membrane system using the same observables as in the validation of the original Martini~3 force field:  area-per-lipid $A_l$, thickness $d$ (both of them obtained with FATSLiM\cite{buchoux2017fatslim}), area compressibility modulus (computed from the projected area fluctuations $K_\mathrm{A}$, without correction for finite size effect\cite{waheed2009undulation}), and $S_n$ order parameter (using \textit{do-order-gmx5.py}, available on the Martini website www.cgmartini.nl). In addition, we also calculated the density profile of the ions along the membrane normal and the lateral diffusion coefficient $D_\mathrm{lat}$ of the lipid centers of mass within the membrane.

\section{Results and Discussion}

\subsection{Stability of Martini~3 NaCl solutions as a function of time step}

\begin{figure}[htb]
\begin{center}
    \includegraphics[width=8.5cm]{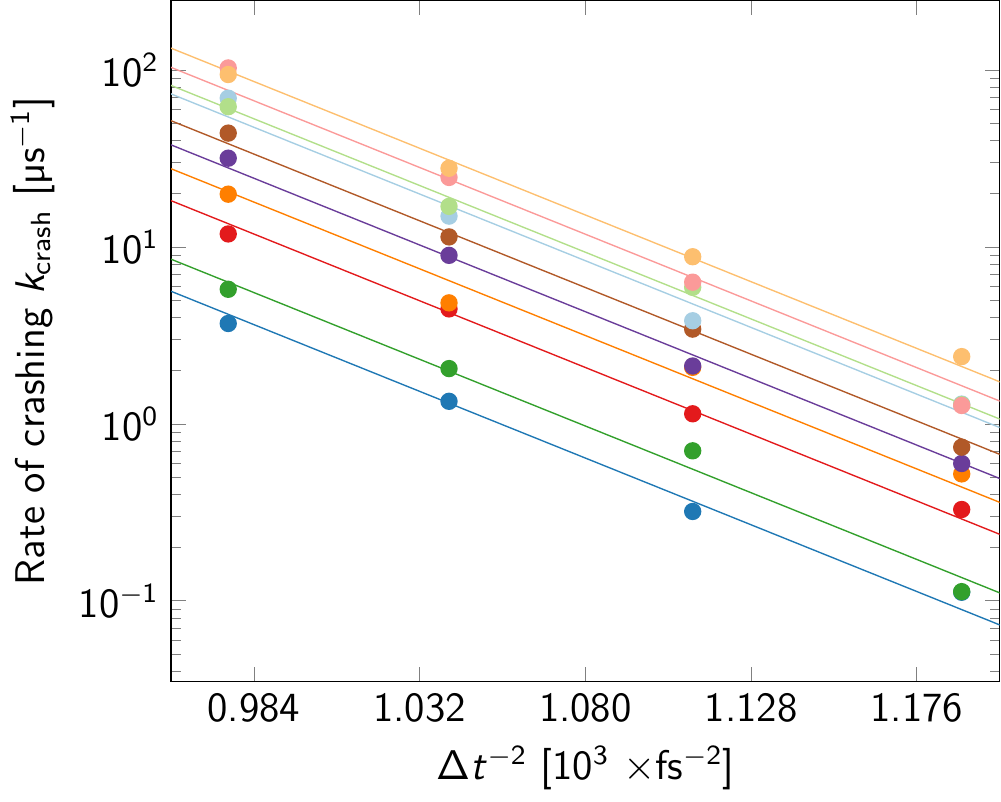}
    \caption{Rate of crashing in the NaCl solutions as a function of the time step size. Shown is $k_{\text{crash}}$ on a semi-logarithmic scale as a function of $1/\Delta t^2$. The individual colors correspond to fixed concentrations. The concentrations increase from the bottom curve (0.075~M) to the top (0.5~M). The solid lines represent fits of Eq.~(\ref{eq:crash-rate}) using a common $\gamma$ value.
    }
    \label{fig:lnk}
\end{center}
\end{figure}

We performed extensive MD simulations of the NaCl solutions at time steps between 20 and 40~fs and a range of ion concentrations. By computing the rate of crashing $k_{\mathrm{crash}}$ using Eq.~(\ref{eq:crash-estimate}), we found more frequent crashes at longer time steps and with increasing number of ions, indicating an issue with the stability of the numerical time integration associated with ions of bead type TQ5. To describe the time step dependence of these crashes, we fitted Eq.~(\ref{eq:crash-rate}) to the $k_{\mathrm{crash}}$ values. As shown in Fig.~\ref{fig:lnk}, the slope of the curves in the semi-logarithmic representation is constant. Accordingly, a single $\gamma$ parameter accurately captures the time-step dependence of the crash tendency independent of ion concentration.

\begin{figure}[htb]
\begin{center}
    \includegraphics[width=8.5cm]{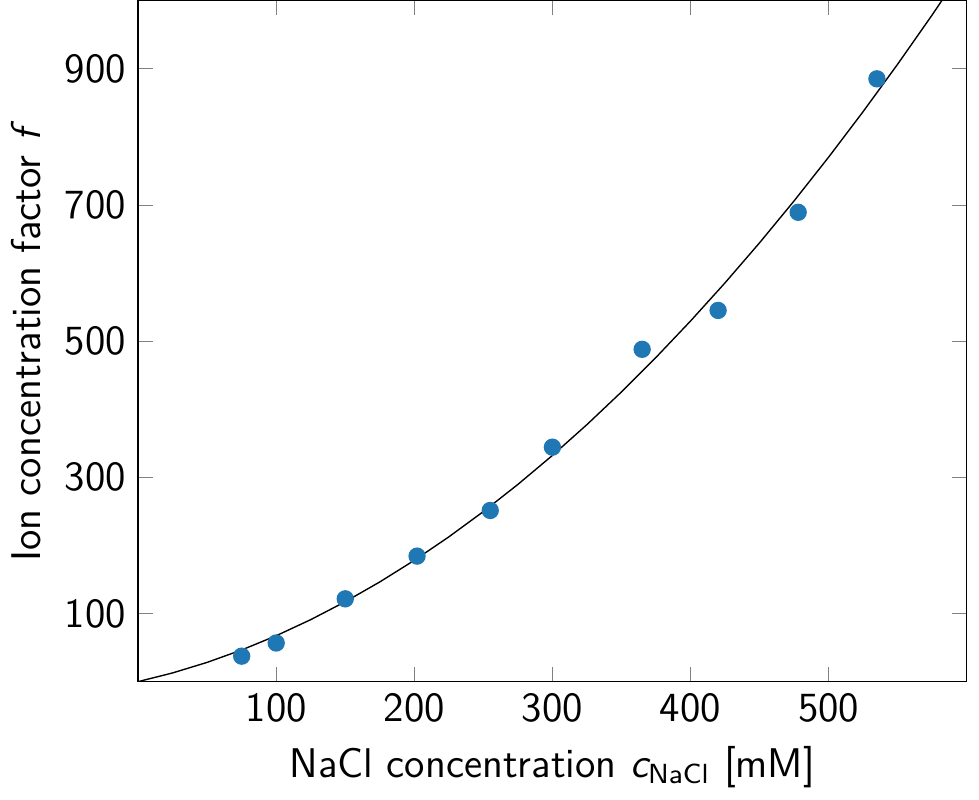}
    \caption{Dependence of rate of crashing on the ion concentration. Shown is the dimensionless concentration factor $f=k_{\mathrm{crash}}^0/\gamma = \left (a c_\mathrm{ion}^2 + b c_\mathrm{ion} \right) V$, $c_\mathrm{ion} = 2 c_\mathrm{NaCl}$ extracted from the intercept of the linear fits in Fig.~\ref{fig:lnk} as a function of the ion concentration in NaCl water solutions. The solid line shows a fit of a linear-quadratic dependence on $c_{\mathrm{ion}}$ with a fixed value of $\gamma$.
    }
    \label{fig:lnk-intercept}
\end{center}
\end{figure}

According to the theory, the prefactor $k_{\mathrm{crash}}^0=f\gamma$ in Eq.~(\ref{eq:crash-rate}) accounts for the ion concentration dependence. Indeed, the values of the dimensionless concentration factor $f=k_{\mathrm{crash}}^0/\gamma$ obtained from the intercept of the fits in Fig.~\ref{fig:lnk} exhibit a linear-quadratic concentration dependence on ion concentration, as shown in Fig.~\ref{fig:lnk-intercept}. We incorporated this concentration dependence and the dependence on the overall system volume $V$ into Eq.~(\ref{eq:crash-rate}) in the form $k_{\mathrm{crash}}^0 \left ( c_\mathrm{ion}, V \right ) = \gamma \left (a c_\mathrm{ion}^2 + b c_\mathrm{ion} \right ) V$. The time step, concentration, and volume dependent crash rate then becomes
\begin{equation}
  \label{eq:crash-rate-full}
  k_\mathrm{crash}\left (c_\mathrm{ion}, V, \Delta t \right) = \gamma \left (a c_\mathrm{ion}^2 + b c_\mathrm{ion} \right ) V \; \exp{\left (-\frac{1}{2\gamma^2 \Delta t^2} \right )},
\end{equation}
where $a = 7.5820 \times 10^{-7}$~mM$^{-2}$~nm$^{-3}$ and $ b = 3.2043 \times 10^{-4}$~mM$^{-1}$~nm$^{-3}$ and $\gamma = 5.2568 $~ps$^{-1}$. Note that a rearrangement of the prefactor gives $k_{\mathrm{crash}}^0 \left ( c_\mathrm{ion}, V \right ) = \gamma n_\mathrm{ions} \left ( a c_\mathrm{ion} + b \right )$, where $ n_\mathrm{ions}/2$ is the total number of positive and negative TQ5 ions in the system.

\begin{figure}[htb]
\begin{center}
    \includegraphics[width=17.0cm]{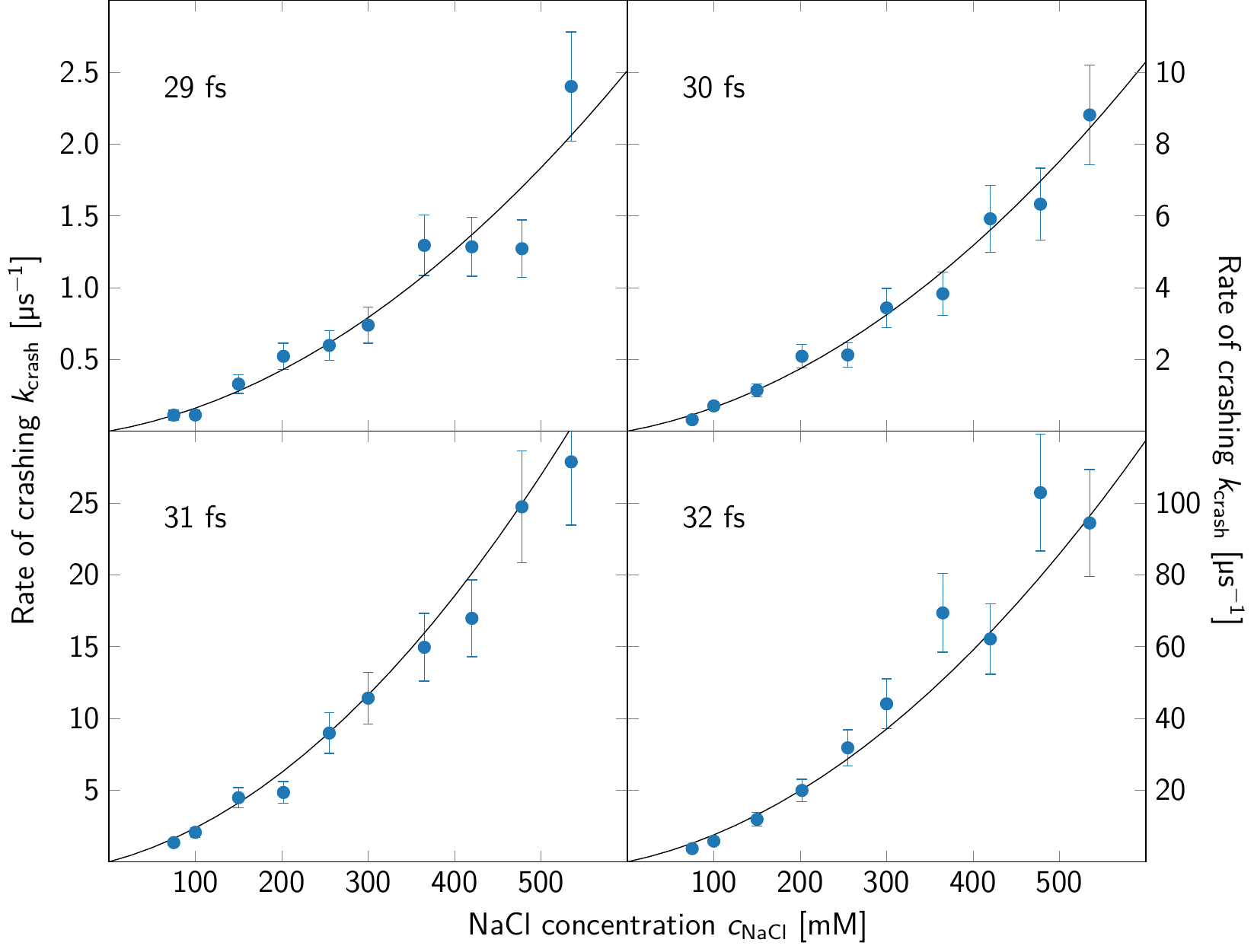}
    \caption{Rate of crashing in the NaCl solutions as a function of NaCl concentration analyzed for different time steps. The panels correspond to $\Delta t = 29,30,31$, and 32~fs. The solid black lines are fits of Eq.~(\ref{eq:crash-estimate}) with three parameters $\gamma$, $a$ and $b$. The error bars represent the standard error. Note that the $y$-scale varies between the panels and that  $c_\mathrm{ion} = 2 c_\mathrm{NaCl}$.
    }
    \label{fig:crash}
\end{center}
\end{figure}

Equation~(\ref{eq:crash-rate}) with a linear-quadratic concentration dependence accurately accounts for the observed crash rates $k_{\mathrm{crash}}$ across ion concentrations and time steps. As shown in Fig.~\ref{fig:crash}, we achieved an excellent agreement between the numerical data for time steps between 29--32 fs and the simple 3-parameter theory. As a final validation, we varied the system size. As predicted by the theory,
$k_{\mathrm{crash}}$ depends linearly on the overall volume of the system (see Fig.~\ref{fig:si-crash-volume}). Incorporating the concentration and system size dependence into Eq.~(\ref{eq:crash-rate}) gives a general equation, Eq.~(\ref{eq:crash-rate-full}), capable of predicting the expected number of crashes. The analysis presented here required a careful control of certain simulation parameters. In particular, we found that using a fixed value of \texttt{verlet-buffer-tolerance} instead of a fixed neighborlist cut-off (\texttt{rlist}) effectively masked the functional form of the concentration dependence of $k_{\mathrm{crash}}$ (see Fig.~\ref{fig:si-crash-rlist}).

\begin{figure}[htb]
\begin{center}
    \includegraphics[width=8.5cm]{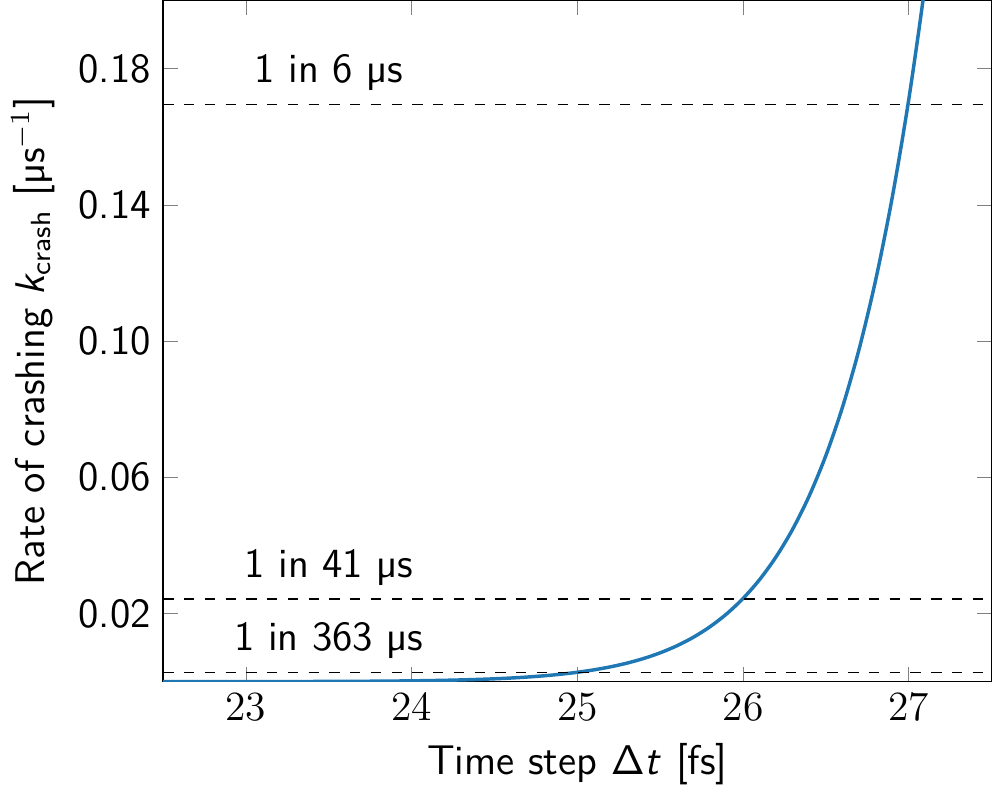}
    \caption{Rate of crashing in a 150~mM NaCl solution consisting of $10^5$ CG beads, as estimated using Eq.~(\ref{eq:crash-rate-full}). Horizontal lines indicate $k_{\text{crash}}$ for runs with time steps of $\Delta t=25$, 26, and 27 fs (bottom to top). The lines are labeled by the expected mean time to a first crash.
    }
    \label{fig:rate-prediction}
\end{center}
\end{figure}

Besides estimating the crash rate, our model also allows to determine the particle types involved in the crashes. At physiological ion concentrations, $c_\mathrm{NaCl} = c_\mathrm{ion}/2 \approx 150$~mM, crashes are more likely to be caused by ion-water collisions than ion-ion collisions. According to the theory, their ratio is about $ac_{\mathrm{ion}}/b\approx 0.71$. Supported by the data, the theory also predicts an extremely steep increase in $k_{\mathrm{crash}}$ as a function of the time step, which was also the main reason for the use of such a seemingly narrow range of $\Delta t$ between 29--32~fs . With relatively short ($10^8$ steps) runs of a small ($\sim$6000 particles) test system, the theory predicts that one would have to perform $\sim$10\ 000~seconds of simulation time at $\Delta t = 20$~fs to reach a single failure even in the most concentrated (500~mM) system. On the other extreme of the scale, $\Delta t = 40$~fs results in a failure after every $\sim$500 steps (every 20~ps), on average. These numbers are supported by our observations in simulations using the respective time steps. As a practical example, we applied Eq.~(\ref{eq:crash-rate-full}) to predict $k_{\mathrm{crash}}$ in a system of 150~mM NaCl solution consisting of $10^5$ particles corresponding to a cubic box of edge length approximately 23~nm. The estimated $k_{\mathrm{crash}}$ as a function of the time step is presented in Fig.~\ref{fig:rate-prediction}. Using $\Delta t = 25$~fs, one can expect one crash about every 366~\textmu s of simulation time, while $\Delta t = 26$~fs already results in a crash every 41 ~\textmu s. We expect that the addition of proteins or a lipid bilayer to the systems does not fundamentally alter $k_{\mathrm{crash}}$, provided they do not introduce additional numerical instabilities. As a consequence, our scaling law limits the largest accessible time step to $\Delta t \leq 25$~fs in any practical application.

\subsection{Increased ionic mass stabilizes time integration}

In numerical tests of the NaCl solutions, we found that an increase in the mass of the ions described by TQ5 beads resulted in more stable time integration. According to Eq.~(\ref{eq:timestep}), we expect that increasing the ionic masses by a factor of $(m'/m)^2=40^2/30^2 \approx 1.7778$ will allow us to use a time step longer by a factor $4/3$. In the more elaborate model Eq.~(\ref{eq:crash-rate}), the ionic mass $m$ enters through $\gamma$. For a modified ionic mass $m'$, we have $\gamma'=\gamma\sqrt{m/m'}$. In practice, the mass $m'$ may have to be adjusted to account for the more complex coupled motions of the different particles. As shown in Fig.~\ref{fig:si-crash-mass}, we obtain excellent agreement of the crash rate predicted  by Eq.~(\ref{eq:crash-rate}) with the data obtained for the heavier ions by using an effective mass of $m'=1.7188 m$ instead of the nominal ionic mass of $1.7778m$, which amounts to a difference of around 3.5{\%}.

These findings further confirm that the ions of TQ5 type indeed determine the stability of the NaCl solutions. Moreover, they also demonstrate that an ionic mass increase improves the stability by allowing a time step scaled approximately by the square root of ionic masses after and before mass correction, in close agreement with the prediction of Eq.~(\ref{eq:timestep}). Taking the 0.3~M test system as an example, doubling the masses at $\Delta t =30$~fs reduces the estimated crash rate $k_{\mathrm{crash}}$ from $\approx 10 / \mbox{\textmu s}$ to $\approx 10^{-8} / \mbox{\textmu s}$, while quadrupling the masses at $\Delta t =40$~fs reduces the estimated $k_{\mathrm{crash}}$ from $\approx 10^5 / \mbox{\textmu s}$ to $\approx 10^{-10} / \mbox{\textmu s}$.

\subsection{Increased ionic masses leave the structure of NaCl solutions unchanged and minimally impact their dynamics}

To investigate the perturbations caused by an increase of ionic masses, we analyzed the structure and dynamics of the NaCl solutions. As expected, the modification of the ionic masses does not impact the structural properties (Fig.~\ref{fig:si-bulk-rdf}). We found the ion-ion radial distribution functions (RDFs) to be independent of the integration time step and the ionic mass. The diffusion coefficients $D_\mathrm{3D}$ of CG water obtained from these systems are collected in Table~\ref{tab:D3d}. The data confirm our expectations based on hydrodynamic considerations \cite{einstein1905molekularkinetischen, saffman1975brownian}: changing the mass of the ions has only small effects on the diffusion coefficient. At a NaCl concentration of 300 mM, a four-fold increase in ionic mass reduced the water diffusivity by about 2{\%}. We expect that ionic mass will impact the librational motion; however, these non-diffusive motions occur on timescales below the frequency of sample collection (Fig.~\ref{fig:si-solvent-gls}). The observed decrease in diffusivity with increasing NaCl concentration is in agreement with the tendency observed in experiments\cite{ben2013influence}. However, we also noticed a mass-independent small decrease in $D_\mathrm{3D}$ with increasing time step $\Delta t$. When compared to the conventional approach of fitting a straight line to an \textit{ad hoc} interval of the MSD curve, the $D_\mathrm{3D}$ values calculated with a GLS estimator show smaller errors, more consistent estimates between replicas and clearer tendencies across the simulations (see Table~\ref{tab:si-D3d} along with Figs.~\ref{fig:si-solvent-gls} and~\ref{fig:si-dcoeff-estimator} of the Supporting Information). Therefore, care must be taken not only when one compares diffusion coefficients from simulations performed using different time steps, but also when using different estimators \cite{bullerjahn2020optimal}.
\begin{table}[]
\centering
\caption{Estimated $D_\mathrm{3D}$ diffusion coefficient [nm$^2$/ns] of the CG water beads in Martini~3 NaCl solutions at various ion concentrations. Results are listed for ionic masses multiplied by factors 1, 2, and 4; and for time steps of 20, 30, and 40 fs. The errors are smaller than 0.01 nm$^2$/ns. These values were obtained from the correctly unwrapped trajectories \cite{buelow2020systematic} using DiffusionGLS \cite{bullerjahn2020optimal,buelow-diffusiongls}. Entries are missing where simulations suffered from a high rate of crashing $k_\mathrm{crash}$. Results are listed without corrections for finite system size\cite{yeh2004system}.}
\label{tab:D3d}
\begin{tabular}{|c|c|ccc|ccc|ccc|}
 \hlineB{4}
\multicolumn{1}{V{4}c}{} & \multicolumn{1}{V{4}c}{0.0~M} & \multicolumn{3}{V{4}cV{4}}{0.075~M}  & \multicolumn{3}{cV{4}}{0.15~M}  & \multicolumn{3}{cV{4}}{0.3~M}  \\ \cline{2-11}
\multicolumn{1}{V{4}c}{} & \multicolumn{1}{V{4}c}{1$\times$m} & \multicolumn{1}{V{4}c|}{1$\times$m} & \multicolumn{1}{c|}{2$\times$m} & \multicolumn{1}{cV{4}}{4$\times$m} & \multicolumn{1}{c|}{1$\times$m}  & \multicolumn{1}{c|}{2$\times$m} & \multicolumn{1}{cV{4}}{4$\times$m} & \multicolumn{1}{c|}{1$\times$m} & \multicolumn{1}{c|}{2$\times$m} & \multicolumn{1}{cV{4}}{4$\times$m} \\ \hlineB{4}
\multicolumn{1}{V{4}c}{20 fs} & \multicolumn{1}{V{4}c}{2.43} & \multicolumn{1}{V{4}c|}{2.26}  & \multicolumn{1}{c|}{2.25}  & \multicolumn{1}{cV{4}}{2.25}  & \multicolumn{1}{c|}{2.11}  & \multicolumn{1}{c|}{2.10}  & \multicolumn{1}{cV{4}}{2.08}  & \multicolumn{1}{c|}{1.81}  & \multicolumn{1}{c|}{1.80}  & \multicolumn{1}{cV{4}}{1.77}  \\ \hline
\multicolumn{1}{V{4}c}{30 fs} & \multicolumn{1}{V{4}c}{2.28} & \multicolumn{1}{V{4}c|}{2.15$^a$}  & \multicolumn{1}{c|}{2.14}  & \multicolumn{1}{cV{4}}{2.14}  & \multicolumn{1}{c|}{---}  & \multicolumn{1}{c|}{2.00}  & \multicolumn{1}{cV{4}}{2.00}  & \multicolumn{1}{c|}{---}  & \multicolumn{1}{c|}{1.73}  & \multicolumn{1}{cV{4}}{1.71}  \\ \hline
\multicolumn{1}{V{4}c}{40 fs} & \multicolumn{1}{V{4}c}{2.10} & \multicolumn{1}{V{4}c|}{---}  & \multicolumn{1}{c|}{1.99$^a$}  & \multicolumn{1}{cV{4}}{1.98}  & \multicolumn{1}{c|}{---}  & \multicolumn{1}{c|}{---}  & \multicolumn{1}{cV{4}}{1.86}  & \multicolumn{1}{c|}{---}  & \multicolumn{1}{c|}{---}  & \multicolumn{1}{cV{4}}{1.62}  \\ \hlineB{4}

\end{tabular}

\footnotesize{$^a$ These simulations had $k_\mathrm{crash}$ around $\mathcal{O} \left ( 1 / \mathrm{simulation}\right)$ }\\
\end{table}

\subsection{Ionic mass increase has no discernible effect on the structure and dynamics in lipid membrane systems}

Having established the lack of appreciable effects of ionic mass in bulk aqueous solution, we turned our attention to the neat DPPC bilayers. As for NaCl solutions, missing entries in all subsequent tables indicate that a high crash rate made it impossible to obtain converged results. The average values of area-per-lipid $A_l$ and membrane thickness $d$ are collected in Tables~\ref{tab:apl} and~\ref{tab:thick}. DPPC lipids modelled with the standard Martini~3 model (denoted M3(-TQ5)) have a smaller $A_l$ and correspondingly higher $d$ than in Martini~2 (M2), consistent with its parametrization. Doubling the mass of TQ5 (system M3-2$\times$m) and subsequently increasing $\Delta t$ has no detectable effect on either of these quantities.

\begin{table}[]
\centering
\caption{Average area-per-lipid $A_l$ [\angstrom$^2$] from simulations of DPPC bilayers. The error is less than $\pm 0.5$~$\angstrom^2$. Entries are missing where simulations failed to run properly. M2 and M3 refer to Martini 2 and Martini 3, respectively.}
\label{tab:apl}
\begin{tabular}{V{4}cV{4}cV{4}cV{4}cV{4}cV{4}cV{4}}
\hlineB{4}
      &  M2  & M3(-TQ5)   & M3-SQ5  & M3-RQ5  &    M3-2$\times$m \\ \hlineB{4}
20 fs & 60.6 &  59.7      &  59.7   &  59.8   &    59.7          \\ \hline
30 fs & 60.5 &   ---      &  59.5   &  59.6   &    59.5          \\ \hline
40 fs & 60.4 &   ---      &    ---  &  59.4   &    ---           \\ \hlineB{4}
\end{tabular}
\end{table}

\begin{table}[]
\centering
\caption{Membrane thickness $d$ [nm] from simulations of DPPC bilayers. The error is less than 0.025~nm. Entries are missing where simulations failed to run properly..}
\label{tab:thick}
\begin{tabular}{V{4}cV{4}cV{4}cV{4}cV{4}cV{4}cV{4}}
\hlineB{4}
      &    M2    &   M3(-TQ5)     &    M3-SQ5    &   M3-RQ5     &    M3-2$\times$m    \\ \hlineB{4}
20 fs & 4.147    &   4.195        &     4.194    &   4.190      &    4.195           \\ \hline
30 fs & 4.151    &   ---          &     4.202    &   4.198      &    4.203            \\ \hline
40 fs & 4.157    &   ---          &    ---       &   4.209      &    ---              \\ \hlineB{4}
\end{tabular}
\end{table}

The lipid models in Martini~3 are designed to be softer than those in Martini~2 \cite{souza2021martini}. Correspondingly, DPPC lipids in the Martini~3 validation exhibited $K_\mathrm{A}$ values of 232.9$\pm$13.3~mN/m in excellent agreement with the experimentally measured 231$\pm$20.0~mN/m\cite{nagle2000structure}. The $K_\mathrm{A}$ values calculated in this work are collected in Table~\ref{tab:compress}. Our values are also in general agreement with the experimental data, but somewhat below them. The discrepancy is not surprising: $K_\mathrm{A}$ values are both sensitive to the length of the time steps\cite{venable2015mechanical} and to system size \cite{venable2015mechanical, saeedimasine2019role}, consistent with our simulations. In particular, the dependence on system size originates from the difference between real and projected membrane areas due to the undulations of the membrane\cite{waheed2009undulation}. A word of caution is needed at this point: a common practice when simulating large patches of planar membranes is to apply weak harmonic restraints to a quarter of the lipids in one of the two leaflets\cite{ingolfsson2014lipid,vogele2018hydrodynamics}. Such ``pinning'' of a subset of lipids suppresses long wavelength undulations. The suppression of undulations should be taken into account when comparing $K_\mathrm{A}$ from simulations of different sizes. As we saw with $A_l$ and $d$, doubling the ionic masses does not influence $K_\mathrm{A}$. However, doubling $\Delta t$ from 20 to 40 fs resulted in a noticeable decrease (10-20{\%}) of $K_\mathrm{A}$ for the M2 and M3-RQ5 systems, where runs with $\Delta t=40$ fs were stable.

\begin{table}[]
\centering
\caption{Area compressibility modulus $K_\mathrm{A}$ [mN/m] from simulations of DPPC bilayers. The error is less than 2~mN/m. Entries are missing where simulations failed to run properly. Results are listed without corrections for finite size\cite{waheed2009undulation}.}
\label{tab:compress}
\begin{tabular}{V{4}cV{4}cV{4}cV{4}cV{4}cV{4}cV{4}}
\hlineB{4}
      &    M2  &   M3(-TQ5)   &    M3-SQ5    &   M3-RQ5   &    M3-2$\times$m    \\ \hlineB{4}
20 fs & 279    &   220        &     221      &   216      &    218              \\ \hline
30 fs & 254    &   ---        &     197      &   197      &    204              \\ \hline
40 fs & 243    &   ---        &    ---       &   179      &    ---              \\ \hlineB{4}
\end{tabular}
\end{table}

The lipid tail order parameter values $S_n$ presented in Table~\ref{tab:order} show a higher degree of order in Martini~3 than in the previous version. Because $S_n$ is computed as an average over the bonds involving lipid tail beads, the values are unaffected by changes in the surrounding medium. An increase in $\Delta t$ causes a minute increase in $S_n$, which matches our expectation of straighter lipid chains based on the computed membrane thickness values.

\begin{table}[]
\centering
\caption{Average order parameter $S_n$ of the carbon chains from simulations of DPPC bilayers. The error is less than 0.001. Entries are missing where simulations failed to run properly.}
\label{tab:order}
\begin{tabular}{V{4}cV{4}cV{4}cV{4}cV{4}cV{4}cV{4}}
\hlineB{4}
      &    M2    &   M3(-TQ5)   &    M3-SQ5    &   M3-RQ5   &    M3-2$\times$m    \\ \hlineB{4}
20 fs & 0.455    &   0.481      &     0.481    &   0.479      &    0.481          \\ \hline
30 fs & 0.457    &   ---        &     0.485    &   0.483      &    0.485          \\ \hline
40 fs & 0.460    &   ---        &    ---       &   0.488      &    ---            \\ \hlineB{4}
\end{tabular}
\end{table}

\begin{figure}[htb]
\begin{center}
    \includegraphics[width=8.5cm]{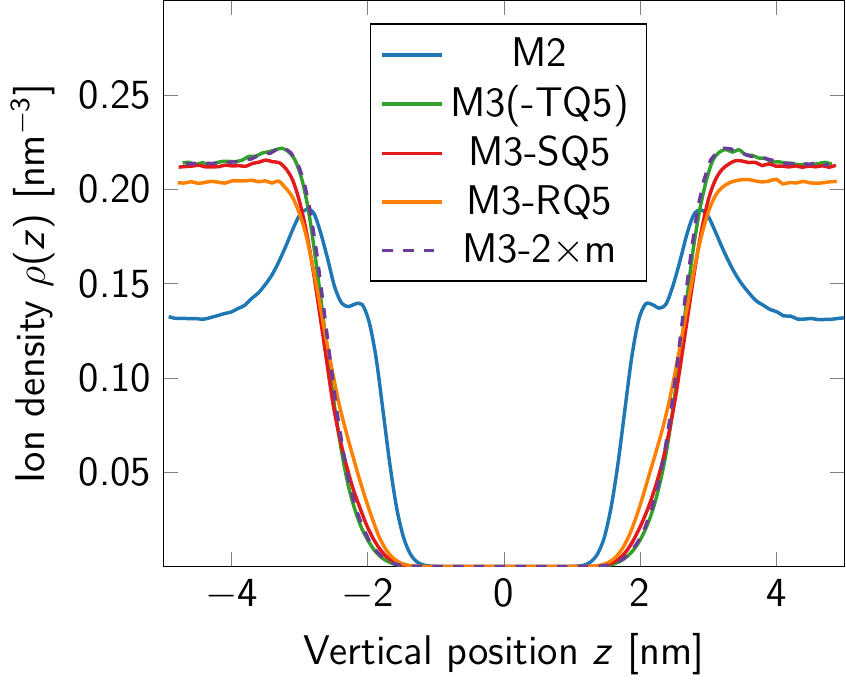}
    \caption{Number density profile of ions along the direction normal to the membrane. The mid-plane of the bilayer is located at $z = 0$~nm. Results are shown for a time step $\Delta t =20$~fs, and are representative of all time steps.
    }
    \label{fig:densprof}
\end{center}
\end{figure}

In addition to the properties used in the optimization of the Martini~3 force field, we calculated the density profile of the ions along the membrane normal and lateral diffusion coefficient $D_\mathrm{lat}$ of the lipid centers of mass. Similar to the radial distribution profiles of the NaCl solutions (Fig.~\ref{fig:si-bulk-rdf}), Fig.~\ref{fig:densprof} demonstrates that the obtained density profiles are insensitive to changes in the ionic mass, which is also true for the separate density profiles of Na$^+$ and Cl$^-$ (see Fig.~\ref{fig:si-membrane-densprof-separate}).

Compared to Martini~2, the lateral diffusion coefficient $D_\mathrm{lat}$ of DPPC lipids is decreased in Martini~3 consistent with the higher order of the lipid chains, as seen from Table~\ref{tab:diffusion}. The GLS estimator again provided diffusion coefficients with smaller statistical errors and clearer tendencies (see Table~\ref{tab:si-Dlat} and Fig.~\ref{fig:si-m3-gls} of the Supporting Information). Note, however, that these diffusion coefficients were not corrected for finite-size effects \cite{vogele2018hydrodynamics}. From the computed $D_\mathrm{lat}$ values it is clear that simply changing the mass of the TQ5 bead without increasing $\Delta t$ does not alter the diffusion coefficient of the lipids, which is in accord with our results for the NaCl solutions.

\begin{table}[]
\centering
\caption{Lateral diffusion coefficients $D_\mathrm{lat}$ [10$^{-2}$ nm$^2$/ns] from simulations of DPPC bilayers. The error is smaller than 0.1$\times$10$^{-2}$ nm$^2$/ns. Entries are missing where simulations failed to run properly.}
\label{tab:diffusion}
\begin{tabular}{V{4}cV{4}cV{4}cV{4}cV{4}cV{4}cV{4}}
\hlineB{4}
      &    M2    &   M3(-TQ5)   &    M3-SQ5    &   M3-RQ5    &    M3-2$\times$m    \\ \hlineB{4}
20 fs & 5.80     &   5.58       &    5.64      &   5.64      &    5.54             \\ \hline
30 fs & 5.69     &   ---        &    5.54      &   5.49      &    5.46             \\ \hline
40 fs & 5.51     &   ---        &    ---       &   5.31      &    ---              \\ \hlineB{4}
\end{tabular}
\end{table}

\subsection{Bead type changes have minimal impact on the structure and dynamics of the membrane}

As an alternative solution to changing the masses, one can modify the bead type used to represent ions in Martini~3. The use of SQ5 beads allowed us to increase $\Delta t$ to 30~fs, meaning not a single crash was detected in the 40 replicas of the NaCl solutions. However, the combination of SQ5 beads and $\Delta t = 40$~fs resulted in an average of 5 crashes per simulation. Further increasing the bead size to RQ5 made it possible to run simulations at $\Delta t = 40$~fs. The procedure of increasing the bead size is fully consistent with the Martini~3 philosophy, and it represents the inclusion of hydrating waters in the ionic bead. Just as in the case of mass changes, we assessed the differences that occur upon changing the bead type from TQ5 to SQ5 or RQ5.

The average values of $A_l$, $d$, and $K_\mathrm{A}$ are listed in Tables~\ref{tab:apl}, \ref{tab:thick}, and~\ref{tab:compress}, respectively, in columns M3-SQ5 and M3-RQ5. Similarly to changing the ionic mass, modifying the bead type while keeping $\Delta t$ constant has virtually no effect on these values. Whereas increasing $\Delta t$ does not affect $A_l$ and $d$ values, we found the $K_\mathrm{A}$ values to decrease noticeably, by 10-20{\%}. Nevertheless, the changes introduced between Martini~2 and Martini~3 are largely preserved independent of the time step. Moreover, the lipid tail order parameters $S_n$ in Table~\ref{tab:order} are insensitive to the ionic bead type, while an increase in $\Delta t$ produces an increase comparable to that observed in the system with doubled masses (also in Table~\ref{tab:order}).

The modification of the bead type altered the distribution of ions around the membrane. However, as Fig.~\ref{fig:densprof} shows, these changes are minor compared to those between the Martini~2 and 3 versions and between different atomistic force fields \cite{catte2016molecular}. The alteration of the bead type also produced a detectable increase in the lipid diffusivity, indicating a somewhat more fluid liquid phase (Table~\ref{tab:diffusion}). However, the effects of changing the time step $\Delta t$ on lipid diffusion are bigger than the effects of changing the ionic mass in the ranges considered.

Finally, to assess the impact of changing the ionic mass or the bead type in systems containing charged lipids, we also tested DPPC bilayers containing 5\% PI-(4,5)P\textsubscript{2}
(phosphatidylinositol 4,5-bisphosphate) \cite{borges2021improved} using the Martini~3 force field. The results are presented in Tables~\ref{tab:si-pip2-apl} to
\ref{tab:si-pip2-dlat} and Figs.~\ref{fig:si-pip2-densprof} and~\ref{fig:si-pip2-densprof-separate}. The changes in the computed quantities were commensurate with those presented here for neat DPPC membranes. However, the PI-(4,5)P\textsubscript{2}-containing simulations were limited to a maximum 30~fs time step.

\section{Conclusions}

The recently developed Martini~3 force field has successfully addressed several fundamental issues raised by \citeauthor{alessandri2019pitfalls}\cite{alessandri2019pitfalls} Most importantly, the introduction of specific cross-interaction terms between particles of different sizes created a consistent framework for the use of different bead sizes.

Here, we showed that the introduction of the ``tiny'' Martini~3 bead type for representing unhydrated Na$^+$ and Cl$^-$ ions limits the accessible time step to below 25~fs. To achieve this, we performed extensive statistical analyses of the rate of crashing and developed a quantitative model of the crash rate $k_{\mathrm{crash}}\left (c_\mathrm{ion}, \Delta t \right)$. The model revealed the role of ion-ion and ion-water collisions in bringing about the crashes. The insight into the factors limiting the stability of MD time integration for neat NaCl aqueous solutions could be transferred to larger and more complex systems containing lipid bilayers. The knowledge of such stability limits greatly facilitates the rational design of computational experiments and the optimal use of available resources. A particular example is the free energy method of \citeauthor{lechner2006equilibrium} relying on fast-switching trajectories, where optimum efficiency of the algorithm was achieved using $\Delta t$ just short of the stability limit. \cite{lechner2006equilibrium, oberhofer2007large} Another example is the setup of in silico high throughput campaigns such as the optimization of compositions of ionic liquids\cite{vazquez2020martini} or deep eutectic solvents\cite{vainikka2021des} for liquid-liquid extractions, where the statistical model can be applied to determine an optimal time step for a given set of molecular systems.

Moreover, our simulations of NaCl solutions demonstrated that increasing the ionic mass has no significant effects on the structure or dynamics of the system (beyond the timescale of librational motions). Doubling (quadrupling) the masses allowed us to perform simulations at $\Delta t = 30$~fs (40~fs), but this resulted in somewhat slower diffusion. The properties of membrane systems containing DPPC lipids were also insensitive to doubling the mass. The larger time step had a minute effect on the structure of the bilayer, and the change in $D_\mathrm{lat}$ was even smaller than in case of the NaCl solutions.

As an alternative consistent with the philosophy of Martini~3, we explored the impact of changing the size of the charged bead used to represent ions. Again, we noticed only a minor influence on the dynamical and structural properties of protein-less bilayers, while making possible the use of an increased time step.

Increased ionic masses or altered bead types allowed us to use time steps of $\Delta t = 30$~fs (with SQ5) or $40$~fs (with RQ5) to model protein-less bilayers. Increased time steps can provide crucial speed-up, e.g., for simulating phase separating lipid mixtures \cite{baoukina2017composition, weiner2018presence}. Although simulations of neat DPPC bilayers were running stably even at 40~fs, we do not recommend going beyond 30~fs. Major reasons to avoid time steps as long as $\Delta t = 40$ fs, besides the general consensus\cite{marrink2010comment}, are the following: (i) the structural and dynamical properties of the system undergo relatively larger changes between 30 and 40~fs than between 20 and 30~fs as seen, e.g., in the diffusion coefficients of water in NaCl solutions; (ii) in the simulations of neat DPPC bilayers, only the M3-RQ5 system could tolerate such a high time step; (iii) the introduction of more finely mapped lipids such as phosphatidylinositol 4,5-bisphosphate\cite{borges2021improved} precludes the use of 40~fs (see Supporting Information). Our analysis indicated that to reach $\Delta t = 30$~fs it suffices to either double the ionic mass or change the bead type to SQ5, which both have only a mild effect on the behavior of the system.

Another known issue that limits the time step in three-component phase separating systems is the presence of insufficiently converged constraints \cite{thallmair2021nonconverged}. We plan to address this problem in a future publication.

\begin{acknowledgement}
B.F. is grateful for Dr. Jakob T\'{o}mas Bullerjahn for fruitful discussions regarding statistical data analysis. This research was supported by the
Max Planck Society (B.F. and G.H). B.F., S.T. and G.H. thank the Center for Multiscale Modelling in Life Sciences (CMMS) sponsored by the Hessian Ministry of Science and Art for funding. S.T. thanks the Alfons und Gertrud Kassel foundation and the Dr. Rolf M. Schwiete foundation.
\end{acknowledgement}

\begin{suppinfo}
See the supplementary material for the definition of crash in simulations, the influence of \texttt{rlist}, doubling of masses, and changing the volume on $k_\mathrm{crash}$, the diffusion coefficients obtained with the conventional estimator and its comparison to the GLS estimator, the ion-ion radial distribution functions in the NaCl solutions, the density profiles of ions around the DPPC membrane, and the results concerning the PI-(4,5)P$_2$ containing membranes.
\end{suppinfo}

\bibliography{refs}

\providecommand{\latin}[1]{#1}
\makeatletter
\providecommand{\doi}
  {\begingroup\let\do\@makeother\dospecials
  \catcode`\{=1 \catcode`\}=2 \doi@aux}
\providecommand{\doi@aux}[1]{\endgroup\texttt{#1}}
\makeatother
\providecommand*\mcitethebibliography{\thebibliography}
\csname @ifundefined\endcsname{endmcitethebibliography}
  {\let\endmcitethebibliography\endthebibliography}{}
\begin{mcitethebibliography}{42}
\providecommand*\natexlab[1]{#1}
\providecommand*\mciteSetBstSublistMode[1]{}
\providecommand*\mciteSetBstMaxWidthForm[2]{}
\providecommand*\mciteBstWouldAddEndPuncttrue
  {\def\EndOfBibitem{\unskip.}}
\providecommand*\mciteBstWouldAddEndPunctfalse
  {\let\EndOfBibitem\relax}
\providecommand*\mciteSetBstMidEndSepPunct[3]{}
\providecommand*\mciteSetBstSublistLabelBeginEnd[3]{}
\providecommand*\EndOfBibitem{}
\mciteSetBstSublistMode{f}
\mciteSetBstMaxWidthForm{subitem}{(\alph{mcitesubitemcount})}
\mciteSetBstSublistLabelBeginEnd
  {\mcitemaxwidthsubitemform\space}
  {\relax}
  {\relax}

\bibitem[Marrink \latin{et~al.}(2004)Marrink, De~Vries, and
  Mark]{marrink2004coarse}
Marrink,~S.~J.; De~Vries,~A.~H.; Mark,~A.~E. Coarse grained model for
  semiquantitative lipid simulations. \emph{The Journal of Physical Chemistry
  B} \textbf{2004}, \emph{108}, 750--760\relax
\mciteBstWouldAddEndPuncttrue
\mciteSetBstMidEndSepPunct{\mcitedefaultmidpunct}
{\mcitedefaultendpunct}{\mcitedefaultseppunct}\relax
\EndOfBibitem
\bibitem[Marrink \latin{et~al.}(2007)Marrink, Risselada, Yefimov, Tieleman, and
  De~Vries]{marrink2007martini}
Marrink,~S.~J.; Risselada,~H.~J.; Yefimov,~S.; Tieleman,~D.~P.; De~Vries,~A.~H.
  The MARTINI force field: coarse grained model for biomolecular simulations.
  \emph{The journal of physical chemistry B} \textbf{2007}, \emph{111},
  7812--7824\relax
\mciteBstWouldAddEndPuncttrue
\mciteSetBstMidEndSepPunct{\mcitedefaultmidpunct}
{\mcitedefaultendpunct}{\mcitedefaultseppunct}\relax
\EndOfBibitem
\bibitem[Souza \latin{et~al.}(2021)Souza, Alessandri, Barnoud, Thallmair,
  Faustino, Gr{\"u}newald, Patmanidis, Abdizadeh, Bruininks, Wassenaar,
  \latin{et~al.} others]{souza2021martini}
Souza,~P.~C.; Alessandri,~R.; Barnoud,~J.; Thallmair,~S.; Faustino,~I.;
  Gr{\"u}newald,~F.; Patmanidis,~I.; Abdizadeh,~H.; Bruininks,~B.~M.;
  Wassenaar,~T.~A., \latin{et~al.}  Martini 3: a general purpose force field
  for coarse-grained molecular dynamics. \emph{Nature methods} \textbf{2021},
  \emph{18}, 382--388\relax
\mciteBstWouldAddEndPuncttrue
\mciteSetBstMidEndSepPunct{\mcitedefaultmidpunct}
{\mcitedefaultendpunct}{\mcitedefaultseppunct}\relax
\EndOfBibitem
\bibitem[Khan \latin{et~al.}(2020)Khan, Souza, Thallmair, Barnoud, De~Vries,
  Marrink, and Reuter]{khan2020capturing}
Khan,~H.~M.; Souza,~P.~C.; Thallmair,~S.; Barnoud,~J.; De~Vries,~A.~H.;
  Marrink,~S.~J.; Reuter,~N. Capturing choline--aromatics Cation- $\pi$
  interactions in the MARTINI force field. \emph{Journal of chemical theory and
  computation} \textbf{2020}, \emph{16}, 2550--2560\relax
\mciteBstWouldAddEndPuncttrue
\mciteSetBstMidEndSepPunct{\mcitedefaultmidpunct}
{\mcitedefaultendpunct}{\mcitedefaultseppunct}\relax
\EndOfBibitem
\bibitem[Weiner and Feigenson(2018)Weiner, and Feigenson]{weiner2018presence}
Weiner,~M.~D.; Feigenson,~G.~W. Presence and role of midplane cholesterol in
  lipid bilayers containing registered or antiregistered phase domains.
  \emph{The Journal of Physical Chemistry B} \textbf{2018}, \emph{122},
  8193--8200\relax
\mciteBstWouldAddEndPuncttrue
\mciteSetBstMidEndSepPunct{\mcitedefaultmidpunct}
{\mcitedefaultendpunct}{\mcitedefaultseppunct}\relax
\EndOfBibitem
\bibitem[Rosetti \latin{et~al.}(2017)Rosetti, Montich, and
  Pastorino]{rosetti2017molecular}
Rosetti,~C.~M.; Montich,~G.~G.; Pastorino,~C. Molecular insight into the line
  tension of bilayer membranes containing hybrid polyunsaturated lipids.
  \emph{The Journal of Physical Chemistry B} \textbf{2017}, \emph{121},
  1587--1600\relax
\mciteBstWouldAddEndPuncttrue
\mciteSetBstMidEndSepPunct{\mcitedefaultmidpunct}
{\mcitedefaultendpunct}{\mcitedefaultseppunct}\relax
\EndOfBibitem
\bibitem[Perlmutter and Sachs(2011)Perlmutter, and
  Sachs]{perlmutter2011interleaflet}
Perlmutter,~J.~D.; Sachs,~J.~N. Interleaflet interaction and asymmetry in phase
  separated lipid bilayers: molecular dynamics simulations. \emph{Journal of
  the American Chemical Society} \textbf{2011}, \emph{133}, 6563--6577\relax
\mciteBstWouldAddEndPuncttrue
\mciteSetBstMidEndSepPunct{\mcitedefaultmidpunct}
{\mcitedefaultendpunct}{\mcitedefaultseppunct}\relax
\EndOfBibitem
\bibitem[De~Jong \latin{et~al.}(2016)De~Jong, Baoukina, Ing{\'o}lfsson, and
  Marrink]{de2016martini}
De~Jong,~D.~H.; Baoukina,~S.; Ing{\'o}lfsson,~H.~I.; Marrink,~S.~J. Martini
  straight: Boosting performance using a shorter cutoff and GPUs.
  \emph{Computer Physics Communications} \textbf{2016}, \emph{199}, 1--7\relax
\mciteBstWouldAddEndPuncttrue
\mciteSetBstMidEndSepPunct{\mcitedefaultmidpunct}
{\mcitedefaultendpunct}{\mcitedefaultseppunct}\relax
\EndOfBibitem
\bibitem[Xing \latin{et~al.}(2009)Xing, Ollila, Vattulainen, and
  Faller]{xing2009asymmetric}
Xing,~C.; Ollila,~O.~S.; Vattulainen,~I.; Faller,~R. Asymmetric nature of
  lateral pressure profiles in supported lipid membranes and its implications
  for membrane protein functions. \emph{Soft Matter} \textbf{2009}, \emph{5},
  3258--3261\relax
\mciteBstWouldAddEndPuncttrue
\mciteSetBstMidEndSepPunct{\mcitedefaultmidpunct}
{\mcitedefaultendpunct}{\mcitedefaultseppunct}\relax
\EndOfBibitem
\bibitem[Jalili and Akhavan(2009)Jalili, and Akhavan]{jalili2009coarse}
Jalili,~S.; Akhavan,~M. A coarse-grained molecular dynamics simulation of a
  sodium dodecyl sulfate micelle in aqueous solution. \emph{Colloids and
  Surfaces A: Physicochemical and Engineering Aspects} \textbf{2009},
  \emph{352}, 99--102\relax
\mciteBstWouldAddEndPuncttrue
\mciteSetBstMidEndSepPunct{\mcitedefaultmidpunct}
{\mcitedefaultendpunct}{\mcitedefaultseppunct}\relax
\EndOfBibitem
\bibitem[Marrink \latin{et~al.}(2010)Marrink, Periole, Tieleman, and
  de~Vries]{marrink2010comment}
Marrink,~S.~J.; Periole,~X.; Tieleman,~D.~P.; de~Vries,~A.~H. Comment on “On
  using a too large integration time step in molecular dynamics simulations of
  coarse-grained molecular models” by M. Winger, D. Trzesniak, R. Baron and
  WF van Gunsteren, Phys. Chem. Chem. Phys., 2009, 11, 1934. \emph{Physical
  Chemistry Chemical Physics} \textbf{2010}, \emph{12}, 2254--2256\relax
\mciteBstWouldAddEndPuncttrue
\mciteSetBstMidEndSepPunct{\mcitedefaultmidpunct}
{\mcitedefaultendpunct}{\mcitedefaultseppunct}\relax
\EndOfBibitem
\bibitem[Lechner \latin{et~al.}(2006)Lechner, Oberhofer, Dellago, and
  Geissler]{lechner2006equilibrium}
Lechner,~W.; Oberhofer,~H.; Dellago,~C.; Geissler,~P.~L. Equilibrium free
  energies from fast-switching trajectories with large time steps. \emph{The
  Journal of chemical physics} \textbf{2006}, \emph{124}, 044113\relax
\mciteBstWouldAddEndPuncttrue
\mciteSetBstMidEndSepPunct{\mcitedefaultmidpunct}
{\mcitedefaultendpunct}{\mcitedefaultseppunct}\relax
\EndOfBibitem
\bibitem[Einstein(1905)]{einstein1905molekularkinetischen}
Einstein,~A. {\"U}ber die von der molekularkinetischen Theorie der W{\"a}rme
  geforderte Bewegung von in ruhenden Fl{\"u}ssigkeiten suspendierten Teilchen.
  \emph{Annalen der physik} \textbf{1905}, \emph{4}\relax
\mciteBstWouldAddEndPuncttrue
\mciteSetBstMidEndSepPunct{\mcitedefaultmidpunct}
{\mcitedefaultendpunct}{\mcitedefaultseppunct}\relax
\EndOfBibitem
\bibitem[Saffman and Delbr{\"u}ck(1975)Saffman, and
  Delbr{\"u}ck]{saffman1975brownian}
Saffman,~P.; Delbr{\"u}ck,~M. Brownian Motion in Biological Membranes.
  \emph{Proc. Natl. Acad. Sci. USA} \textbf{1975}, \emph{72}, 3111--3113\relax
\mciteBstWouldAddEndPuncttrue
\mciteSetBstMidEndSepPunct{\mcitedefaultmidpunct}
{\mcitedefaultendpunct}{\mcitedefaultseppunct}\relax
\EndOfBibitem
\bibitem[Yeh and Hummer(2004)Yeh, and Hummer]{yeh2004system}
Yeh,~I.-C.; Hummer,~G. System-size dependence of diffusion coefficients and
  viscosities from molecular dynamics simulations with periodic boundary
  conditions. \emph{The Journal of Physical Chemistry B} \textbf{2004},
  \emph{108}, 15873--15879\relax
\mciteBstWouldAddEndPuncttrue
\mciteSetBstMidEndSepPunct{\mcitedefaultmidpunct}
{\mcitedefaultendpunct}{\mcitedefaultseppunct}\relax
\EndOfBibitem
\bibitem[V{\"o}gele \latin{et~al.}(2018)V{\"o}gele, K{\"o}finger, and
  Hummer]{vogele2018hydrodynamics}
V{\"o}gele,~M.; K{\"o}finger,~J.; Hummer,~G. Hydrodynamics of diffusion in
  lipid membrane simulations. \emph{Physical review letters} \textbf{2018},
  \emph{120}, 268104\relax
\mciteBstWouldAddEndPuncttrue
\mciteSetBstMidEndSepPunct{\mcitedefaultmidpunct}
{\mcitedefaultendpunct}{\mcitedefaultseppunct}\relax
\EndOfBibitem
\bibitem[Wassenaar \latin{et~al.}(2015)Wassenaar, Ing{\'o}lfsson, B\"{o}ckmann,
  Tieleman, and Marrink]{wassenaar2015computational}
Wassenaar,~T.~A.; Ing{\'o}lfsson,~H.~I.; B\"{o}ckmann,~R.~A.; Tieleman,~D.~P.;
  Marrink,~S.~J. Computational lipidomics with insane: a versatile tool for
  generating custom membranes for molecular simulations. \emph{Journal of
  chemical theory and computation} \textbf{2015}, \emph{11}, 2144--2155\relax
\mciteBstWouldAddEndPuncttrue
\mciteSetBstMidEndSepPunct{\mcitedefaultmidpunct}
{\mcitedefaultendpunct}{\mcitedefaultseppunct}\relax
\EndOfBibitem
\bibitem[Abraham \latin{et~al.}(2015)Abraham, Murtola, Schulz, P{\'a}ll, Smith,
  Hess, and Lindahl]{abraham2015gromacs}
Abraham,~M.~J.; Murtola,~T.; Schulz,~R.; P{\'a}ll,~S.; Smith,~J.~C.; Hess,~B.;
  Lindahl,~E. GROMACS: High performance molecular simulations through
  multi-level parallelism from laptops to supercomputers. \emph{SoftwareX}
  \textbf{2015}, \emph{1}, 19--25\relax
\mciteBstWouldAddEndPuncttrue
\mciteSetBstMidEndSepPunct{\mcitedefaultmidpunct}
{\mcitedefaultendpunct}{\mcitedefaultseppunct}\relax
\EndOfBibitem
\bibitem[Bussi \latin{et~al.}(2007)Bussi, Donadio, and
  Parrinello]{bussi2007canonical}
Bussi,~G.; Donadio,~D.; Parrinello,~M. Canonical sampling through velocity
  rescaling. \emph{The Journal of chemical physics} \textbf{2007}, \emph{126},
  014101\relax
\mciteBstWouldAddEndPuncttrue
\mciteSetBstMidEndSepPunct{\mcitedefaultmidpunct}
{\mcitedefaultendpunct}{\mcitedefaultseppunct}\relax
\EndOfBibitem
\bibitem[Parrinello and Rahman(1981)Parrinello, and
  Rahman]{parrinello1981polymorphic}
Parrinello,~M.; Rahman,~A. Polymorphic transitions in single crystals: A new
  molecular dynamics method. \emph{Journal of Applied physics} \textbf{1981},
  \emph{52}, 7182--7190\relax
\mciteBstWouldAddEndPuncttrue
\mciteSetBstMidEndSepPunct{\mcitedefaultmidpunct}
{\mcitedefaultendpunct}{\mcitedefaultseppunct}\relax
\EndOfBibitem
\bibitem[Humphrey \latin{et~al.}(1996)Humphrey, Dalke, and Schulten]{HUMP96}
Humphrey,~W.; Dalke,~A.; Schulten,~K. {VMD} -- {V}isual {M}olecular {D}ynamics.
  \emph{Journal of Molecular Graphics} \textbf{1996}, \emph{14}, 33--38\relax
\mciteBstWouldAddEndPuncttrue
\mciteSetBstMidEndSepPunct{\mcitedefaultmidpunct}
{\mcitedefaultendpunct}{\mcitedefaultseppunct}\relax
\EndOfBibitem
\bibitem[Berendsen \latin{et~al.}(1984)Berendsen, Postma, van Gunsteren,
  DiNola, and Haak]{berendsen1984molecular}
Berendsen,~H.~J.; Postma,~J.~v.; van Gunsteren,~W.~F.; DiNola,~A.; Haak,~J.~R.
  Molecular dynamics with coupling to an external bath. \emph{The Journal of
  chemical physics} \textbf{1984}, \emph{81}, 3684--3690\relax
\mciteBstWouldAddEndPuncttrue
\mciteSetBstMidEndSepPunct{\mcitedefaultmidpunct}
{\mcitedefaultendpunct}{\mcitedefaultseppunct}\relax
\EndOfBibitem
\bibitem[Bullerjahn \latin{et~al.}(2020)Bullerjahn, von B{\"u}low, and
  Hummer]{bullerjahn2020optimal}
Bullerjahn,~J.~T.; von B{\"u}low,~S.; Hummer,~G. Optimal estimates of
  self-diffusion coefficients from molecular dynamics simulations. \emph{The
  Journal of Chemical Physics} \textbf{2020}, \emph{153}, 024116\relax
\mciteBstWouldAddEndPuncttrue
\mciteSetBstMidEndSepPunct{\mcitedefaultmidpunct}
{\mcitedefaultendpunct}{\mcitedefaultseppunct}\relax
\EndOfBibitem
\bibitem[bue()]{buelow-diffusiongls}
Diffusion Coefficient Fitting.
  \url{https://github.com/bio-phys/DiffusionGLS}\relax
\mciteBstWouldAddEndPuncttrue
\mciteSetBstMidEndSepPunct{\mcitedefaultmidpunct}
{\mcitedefaultendpunct}{\mcitedefaultseppunct}\relax
\EndOfBibitem
\bibitem[von B{\"u}low \latin{et~al.}(2020)von B{\"u}low, Bullerjahn, and
  Hummer]{buelow2020systematic}
von B{\"u}low,~S.; Bullerjahn,~J.~T.; Hummer,~G. Systematic errors in diffusion
  coefficients from long-time molecular dynamics simulations at constant
  pressure. \emph{The Journal of Chemical Physics} \textbf{2020}, \emph{153},
  021101\relax
\mciteBstWouldAddEndPuncttrue
\mciteSetBstMidEndSepPunct{\mcitedefaultmidpunct}
{\mcitedefaultendpunct}{\mcitedefaultseppunct}\relax
\EndOfBibitem
\bibitem[hen()]{henine-qwrap}
qwrap — Fast PBC wrapping and unwrapping for VMD.
  \url{https://github.com/jhenin/qwrap}\relax
\mciteBstWouldAddEndPuncttrue
\mciteSetBstMidEndSepPunct{\mcitedefaultmidpunct}
{\mcitedefaultendpunct}{\mcitedefaultseppunct}\relax
\EndOfBibitem
\bibitem[Buchoux(2017)]{buchoux2017fatslim}
Buchoux,~S. FATSLiM: a fast and robust software to analyze MD simulations of
  membranes. \emph{Bioinformatics} \textbf{2017}, \emph{33}, 133--134\relax
\mciteBstWouldAddEndPuncttrue
\mciteSetBstMidEndSepPunct{\mcitedefaultmidpunct}
{\mcitedefaultendpunct}{\mcitedefaultseppunct}\relax
\EndOfBibitem
\bibitem[Waheed and Edholm(2009)Waheed, and Edholm]{waheed2009undulation}
Waheed,~Q.; Edholm,~O. Undulation contributions to the area compressibility in
  lipid bilayer simulations. \emph{Biophysical journal} \textbf{2009},
  \emph{97}, 2754--2760\relax
\mciteBstWouldAddEndPuncttrue
\mciteSetBstMidEndSepPunct{\mcitedefaultmidpunct}
{\mcitedefaultendpunct}{\mcitedefaultseppunct}\relax
\EndOfBibitem
\bibitem[Ben~Ishai \latin{et~al.}(2013)Ben~Ishai, Mamontov, Nickels, and
  Sokolov]{ben2013influence}
Ben~Ishai,~P.; Mamontov,~E.; Nickels,~J.~D.; Sokolov,~A.~P. Influence of Ions
  on Water Diffusion---A Neutron Scattering Study. \emph{The Journal of
  Physical Chemistry B} \textbf{2013}, \emph{117}, 7724--7728\relax
\mciteBstWouldAddEndPuncttrue
\mciteSetBstMidEndSepPunct{\mcitedefaultmidpunct}
{\mcitedefaultendpunct}{\mcitedefaultseppunct}\relax
\EndOfBibitem
\bibitem[Nagle and Tristram-Nagle(2000)Nagle, and
  Tristram-Nagle]{nagle2000structure}
Nagle,~J.~F.; Tristram-Nagle,~S. Structure of lipid bilayers. \emph{Biochimica
  et Biophysica Acta (BBA)-Reviews on Biomembranes} \textbf{2000}, \emph{1469},
  159--195\relax
\mciteBstWouldAddEndPuncttrue
\mciteSetBstMidEndSepPunct{\mcitedefaultmidpunct}
{\mcitedefaultendpunct}{\mcitedefaultseppunct}\relax
\EndOfBibitem
\bibitem[Venable \latin{et~al.}(2015)Venable, Brown, and
  Pastor]{venable2015mechanical}
Venable,~R.~M.; Brown,~F.~L.; Pastor,~R.~W. Mechanical properties of lipid
  bilayers from molecular dynamics simulation. \emph{Chemistry and physics of
  lipids} \textbf{2015}, \emph{192}, 60--74\relax
\mciteBstWouldAddEndPuncttrue
\mciteSetBstMidEndSepPunct{\mcitedefaultmidpunct}
{\mcitedefaultendpunct}{\mcitedefaultseppunct}\relax
\EndOfBibitem
\bibitem[Saeedimasine \latin{et~al.}(2019)Saeedimasine, Montanino, Kleiven, and
  Villa]{saeedimasine2019role}
Saeedimasine,~M.; Montanino,~A.; Kleiven,~S.; Villa,~A. Role of lipid
  composition on the structural and mechanical features of axonal membranes: a
  molecular simulation study. \emph{Scientific reports} \textbf{2019},
  \emph{9}, 1--12\relax
\mciteBstWouldAddEndPuncttrue
\mciteSetBstMidEndSepPunct{\mcitedefaultmidpunct}
{\mcitedefaultendpunct}{\mcitedefaultseppunct}\relax
\EndOfBibitem
\bibitem[Ing{\'o}lfsson \latin{et~al.}(2014)Ing{\'o}lfsson, Melo, Van~Eerden,
  Arnarez, Lopez, Wassenaar, Periole, De~Vries, Tieleman, and
  Marrink]{ingolfsson2014lipid}
Ing{\'o}lfsson,~H.~I.; Melo,~M.~N.; Van~Eerden,~F.~J.; Arnarez,~C.;
  Lopez,~C.~A.; Wassenaar,~T.~A.; Periole,~X.; De~Vries,~A.~H.;
  Tieleman,~D.~P.; Marrink,~S.~J. Lipid organization of the plasma membrane.
  \emph{Journal of the american chemical society} \textbf{2014}, \emph{136},
  14554--14559\relax
\mciteBstWouldAddEndPuncttrue
\mciteSetBstMidEndSepPunct{\mcitedefaultmidpunct}
{\mcitedefaultendpunct}{\mcitedefaultseppunct}\relax
\EndOfBibitem
\bibitem[Catte \latin{et~al.}(2016)Catte, Girych, Javanainen, Loison, Melcr,
  Miettinen, Monticelli, M{\"a}{\"a}tt{\"a}, Oganesyan, Ollila, \latin{et~al.}
  others]{catte2016molecular}
Catte,~A.; Girych,~M.; Javanainen,~M.; Loison,~C.; Melcr,~J.; Miettinen,~M.~S.;
  Monticelli,~L.; M{\"a}{\"a}tt{\"a},~J.; Oganesyan,~V.~S.; Ollila,~O.~S.,
  \latin{et~al.}  Molecular electrometer and binding of cations to phospholipid
  bilayers. \emph{Physical Chemistry Chemical Physics} \textbf{2016},
  \emph{18}, 32560--32569\relax
\mciteBstWouldAddEndPuncttrue
\mciteSetBstMidEndSepPunct{\mcitedefaultmidpunct}
{\mcitedefaultendpunct}{\mcitedefaultseppunct}\relax
\EndOfBibitem
\bibitem[Borges-Ara{\'u}jo \latin{et~al.}(2021)Borges-Ara{\'u}jo, Souza,
  Fernandes, and Melo]{borges2021improved}
Borges-Ara{\'u}jo,~L.; Souza,~P.; Fernandes,~F.; Melo,~M.~N. Improved
  parameterization of phosphatidylinositide lipid headgroups for the Martini 3
  coarse grain force field. \textbf{2021}, \relax
\mciteBstWouldAddEndPunctfalse
\mciteSetBstMidEndSepPunct{\mcitedefaultmidpunct}
{}{\mcitedefaultseppunct}\relax
\EndOfBibitem
\bibitem[Alessandri \latin{et~al.}(2019)Alessandri, Souza, Thallmair, Melo,
  De~Vries, and Marrink]{alessandri2019pitfalls}
Alessandri,~R.; Souza,~P.~C.; Thallmair,~S.; Melo,~M.~N.; De~Vries,~A.~H.;
  Marrink,~S.~J. Pitfalls of the Martini model. \emph{Journal of chemical
  theory and computation} \textbf{2019}, \emph{15}, 5448--5460\relax
\mciteBstWouldAddEndPuncttrue
\mciteSetBstMidEndSepPunct{\mcitedefaultmidpunct}
{\mcitedefaultendpunct}{\mcitedefaultseppunct}\relax
\EndOfBibitem
\bibitem[Oberhofer and Dellago(2007)Oberhofer, and Dellago]{oberhofer2007large}
Oberhofer,~H.; Dellago,~C. Large time-step, fast-switching free energy
  calculations with non-symplectic integrators. \emph{Israel Journal of
  Chemistry} \textbf{2007}, \emph{47}, 215--223\relax
\mciteBstWouldAddEndPuncttrue
\mciteSetBstMidEndSepPunct{\mcitedefaultmidpunct}
{\mcitedefaultendpunct}{\mcitedefaultseppunct}\relax
\EndOfBibitem
\bibitem[Vazquez-Salazar \latin{et~al.}(2020)Vazquez-Salazar, Selle, De~Vries,
  Marrink, and Souza]{vazquez2020martini}
Vazquez-Salazar,~L.~I.; Selle,~M.; De~Vries,~A.~H.; Marrink,~S.~J.;
  Souza,~P.~C. Martini coarse-grained models of imidazolium-based ionic
  liquids: from nanostructural organization to liquid--liquid extraction.
  \emph{Green Chemistry} \textbf{2020}, \emph{22}, 7376--7386\relax
\mciteBstWouldAddEndPuncttrue
\mciteSetBstMidEndSepPunct{\mcitedefaultmidpunct}
{\mcitedefaultendpunct}{\mcitedefaultseppunct}\relax
\EndOfBibitem
\bibitem[Vainikka \latin{et~al.}(2021)Vainikka, Thallmair, Souza, and
  Marrink]{vainikka2021des}
Vainikka,~P.; Thallmair,~S.; Souza,~P. C.~T.; Marrink,~S.~J. Martini 3
  Coarse-Grained Model for Type III Deep Eutectic Solvents: Thermodynamic,
  Structural, and Extraction Properties. \emph{ACS Sustainable Chemistry \&
  Engineering} \textbf{2021}, \emph{9}, 17338--17350\relax
\mciteBstWouldAddEndPuncttrue
\mciteSetBstMidEndSepPunct{\mcitedefaultmidpunct}
{\mcitedefaultendpunct}{\mcitedefaultseppunct}\relax
\EndOfBibitem
\bibitem[Baoukina \latin{et~al.}(2017)Baoukina, Rozmanov, and
  Tieleman]{baoukina2017composition}
Baoukina,~S.; Rozmanov,~D.; Tieleman,~D.~P. Composition fluctuations in lipid
  bilayers. \emph{Biophysical journal} \textbf{2017}, \emph{113},
  2750--2761\relax
\mciteBstWouldAddEndPuncttrue
\mciteSetBstMidEndSepPunct{\mcitedefaultmidpunct}
{\mcitedefaultendpunct}{\mcitedefaultseppunct}\relax
\EndOfBibitem
\bibitem[Thallmair \latin{et~al.}(2021)Thallmair, Javanainen, F{\'a}bi{\'a}n,
  Martinez-Seara, and Marrink]{thallmair2021nonconverged}
Thallmair,~S.; Javanainen,~M.; F{\'a}bi{\'a}n,~B.; Martinez-Seara,~H.;
  Marrink,~S.~J. Nonconverged Constraints Cause Artificial Temperature
  Gradients in Lipid Bilayer Simulations. \emph{The Journal of Physical
  Chemistry B} \textbf{2021}, \emph{125}, 9537--9546\relax
\mciteBstWouldAddEndPuncttrue
\mciteSetBstMidEndSepPunct{\mcitedefaultmidpunct}
{\mcitedefaultendpunct}{\mcitedefaultseppunct}\relax
\EndOfBibitem
\end{mcitethebibliography}

\makeatletter\@input{a.tex}\makeatother
\makeatletter\@input{b.tex}\makeatother

\end{document}


\beginsupplement

\newpage
\tableofcontents

\newpage
\section{Gromacs error messages}
Diagnosing issues when using highly optimized MD codes is not a trivial task. With the exception of LINCS-related issues, Gromacs generally does not print out the steps leading to a crash, therefore problems are generally elusive and difficult to track down. Here, we mention the encountered error messages to provide guidance in detecting the above discussed problem.

\subsection{Errors in the solvent simulations}

All solvent simulations were preformed using 1 node with 1 GPU. The 3 types of error observed were the following:
\begin{itemize}
    \item the simulation stalled: the process was active, but nothing was written to the logfile.
    \item the simulation stopped with an error message after several \texttt{Warning: Pressure scaling more than 1\%}.
    \item one update resulted in ``not-a-number'' (NAN), but the simulation continued until the prescribed number of timesteps.
    Most of the logged values and final averages were NAN.
\end{itemize}
We considered a simulation as crashed at the first occurrence of any of these error.

\subsection{Errors in the membrane simulations}
An invalid displacement of the ions can make the interacting partner explode. Such explosions result in the atoms being scattered way outside the simulation box leading to non-physical geometries. In the $NPT$ ensemble, this is usually followed by a huge scaling of the simulation box and a warning from the barostat. Finally, if constraints are present, this leads to LINCS issues \texttt{ LINCS WARNING relative constraint deviation after LINCS ...} and eventual failure \texttt{One or more atoms moved too far between two domain decomposition steps}.

\newpage
\section{Effect of increasing the system volume on $k_\mathrm{crash}$}

\begin{figure}[htb]
\begin{center}
    \includegraphics[width=8.5cm]{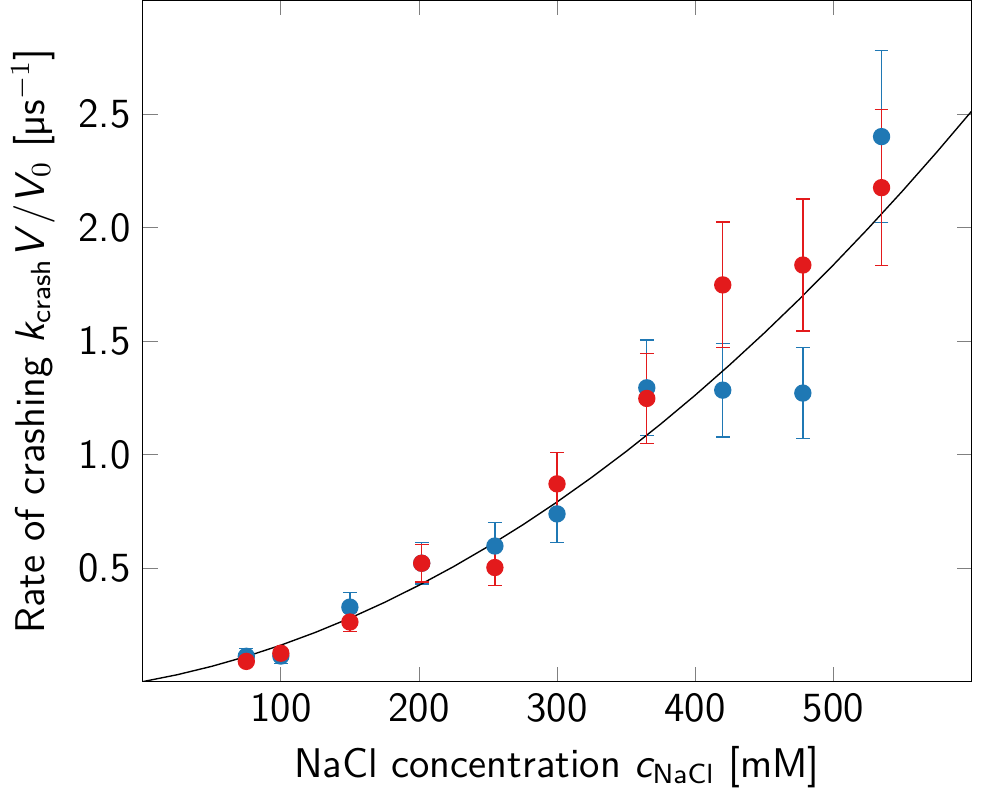}
    \caption{Volume-scaled rate of crashing $k_\mathrm{crash}V_0/V$ in the original solvent simulations of volume $V_0$ (blue markers) and in a system with volume $V=8V_0$ (red markers) at $\Delta t = 29$~fs. Note that $c_\mathrm{ion} = 2 c_\mathrm{NaCl}$.
    }
    \label{fig:si-crash-volume}
\end{center}
\end{figure}

To verify that the rate of crashing $k_\mathrm{crash}$ is an extensive property of the system, we increased the linear system size of the NaCl solution by a factor of 2, scaling the volume by a factor of 8. The $k_\mathrm{crash}$ values computed using $\Delta t= 29$~fs are presented in Fig.~\ref{fig:si-crash-volume}, where the $k_\mathrm{crash}$ values of the larger system were divided by 8 to account for the volume difference.

\newpage
\section{Influence of \texttt{rlist} on $k_\mathrm{crash}$}

\begin{figure}[htb]
\begin{center}
    \includegraphics[width=8.5cm]{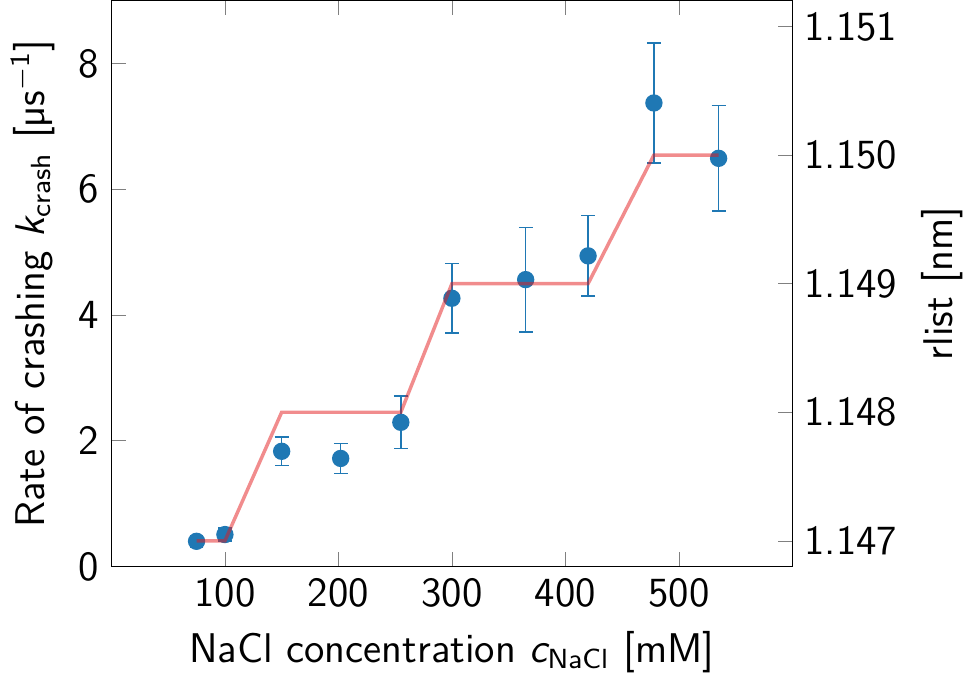}
    \caption{Rate of crashing in the solvent simulations as a function of concentration, at $\Delta t = 30$~fs. The red line corresponds to the \texttt{rlist} simulation parameter, as described in the text.
    }
    \label{fig:si-crash-rlist}
\end{center}
\end{figure}

To study $k_\mathrm{crash}$ as a function of concentration and time step, we kept constant all the other molecular dynamics settings. However, as Fig.~\ref{fig:si-crash-rlist} illustrates, a fixed value of \texttt{verlet-buffer-tolerance} resulted in a step-like change in $k_\mathrm{crash}$ as a function of the concentration. The ``inner list'' value of the ``equivalent classical 1x1 list'', as listed in the log file, exhibited prefect correlation with the
step-like behavior of $k_\mathrm{crash}$. Moreover, fixing the \texttt{rlist} value itself in the mdp file eliminated the step-like changes, and we obtained smoothly varying values of $k_\mathrm{crash}$, as shown in the main text.

\newpage
\section{Effect of increased ion mass on crash rate}

\begin{figure}[htb]
\begin{center}
    \includegraphics[width=8.5cm]{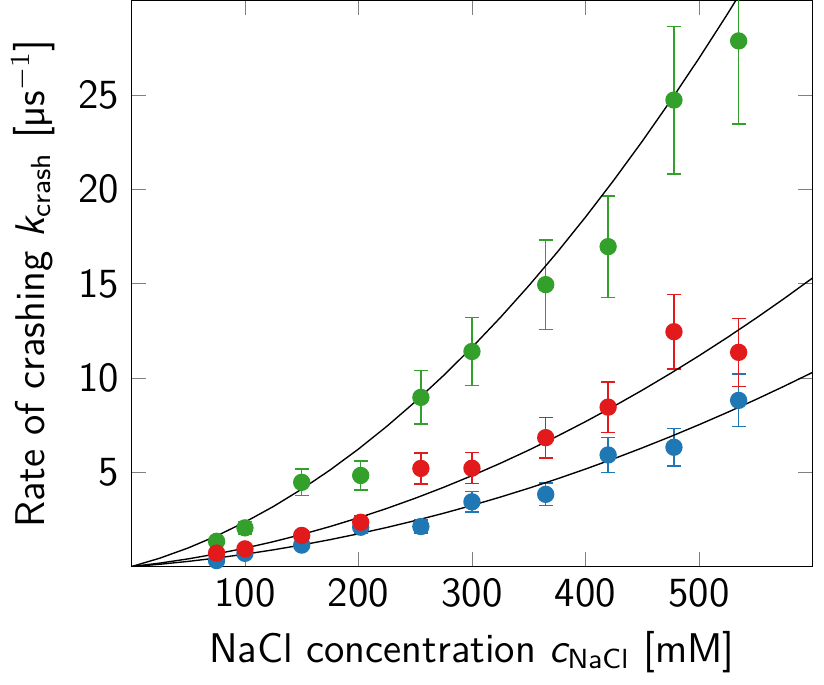}
    \caption{Rate of crashing $k_\mathrm{crash}$ in the solvent simulations using ions with a mass scaled by a factor $40^2/30^2 \approx 1.7778$ and a time step of $\Delta t = 40$~fs (red markers). The fit of Eq. (5) in the main text gave an effective mass increase of $m'=1.7188m$, about 3.5{\%} lower than expected from the ratio of the nominal ion masses. For reference, the blue and green symbols show the values of $k_\mathrm{crash}$ obtained with unmodified ion masses at time steps of 30 and 31~fs, respectively, while the continuous lines are fits of Eq.~\ref{eq:crash-rate-full}.
    }
    \label{fig:si-crash-mass}
\end{center}
\end{figure}

According to Eq.~(\ref{eq:timestep}) in the main text, scaling the masses by $\Delta t'^2 / \Delta t^2$ should reduce $k_\mathrm{crash}$ from those observed at $\Delta t'$ to $\Delta t$. We tested this relationship by increasing the masses of the ions by a factor of $40^2/30^2 \approx 1.7778$, expecting the resulting rate to be similar to the $\Delta t = 30$~fs simulations (see Fig.~\ref{fig:si-crash-mass}). The resulting $k_\mathrm{crash}$ (red markers) was somewhere between that of the runs with unmodified ion masses and time steps of 30 and 31~fs (blue and green). A direct fit of Eq.~(\ref{eq:crash-rate-full}) with variable $\gamma'=\gamma\sqrt{m/m'}$ gave an effective mass of $m'=1.7188m$, which is very close to the nominal ion mass of $1.7778m$. These results indicate that the scaling of the masses indeed closely follows Eq.~(\ref{eq:timestep}).

\newpage
\section{Diffusion coefficients obtained with conventional least squares fitting}

\begin{table}[htb]
\centering
\caption{Estimated $D_\mathrm{3D}$ diffusion coefficient [nm$^2$/ns] of various concentrations of Martini 3 NaCl solutions at increasing masses of the ions and at various timesteps. The error is less than $\pm 0.01$~nm$^2$/ns. Entries are missing where simulations failed to run properly. The listed values were obtained from the unwrapped trajectories \texttt{gmx msd -s topol.tpr -f nojump.xtc -rmcomm -trestart 10000 -o msd-unwrapped.xvg}.}
\label{tab:si-D3d}
\begin{tabular}{|c|c|ccc|ccc|ccc|}
 \hlineB{4}
\multicolumn{1}{V{4}c}{} & \multicolumn{1}{V{4}c}{0.0~M} & \multicolumn{3}{V{4}cV{4}}{0.075~M}  & \multicolumn{3}{cV{4}}{0.15~M}  & \multicolumn{3}{cV{4}}{0.3~M}  \\ \cline{2-11}
\multicolumn{1}{V{4}c}{} & \multicolumn{1}{V{4}c}{1$\times$m} & \multicolumn{1}{V{4}c|}{1$\times$m} & \multicolumn{1}{c|}{2$\times$m} & \multicolumn{1}{cV{4}}{4$\times$m} & \multicolumn{1}{c|}{1$\times$m}  & \multicolumn{1}{c|}{2$\times$m} & \multicolumn{1}{cV{4}}{4$\times$m} & \multicolumn{1}{c|}{1$\times$m} & \multicolumn{1}{c|}{2$\times$m} & \multicolumn{1}{cV{4}}{4$\times$m} \\ \hlineB{4}
\multicolumn{1}{V{4}c}{20 fs} & \multicolumn{1}{V{4}c}{2.44} & \multicolumn{1}{V{4}c|}{2.27}  & \multicolumn{1}{c|}{2.29}  & \multicolumn{1}{cV{4}}{2.23}  & \multicolumn{1}{c|}{2.08}  & \multicolumn{1}{c|}{2.13}  & \multicolumn{1}{cV{4}}{2.08}  & \multicolumn{1}{c|}{1.82}  & \multicolumn{1}{c|}{1.74}  & \multicolumn{1}{cV{4}}{1.74}  \\ \hline
\multicolumn{1}{V{4}c}{30 fs} & \multicolumn{1}{V{4}c}{2.28} & \multicolumn{1}{V{4}c|}{2.16}  & \multicolumn{1}{c|}{2.17}  & \multicolumn{1}{cV{4}}{2.13}  & \multicolumn{1}{c|}{---}  & \multicolumn{1}{c|}{2.00}  & \multicolumn{1}{cV{4}}{1.99}  & \multicolumn{1}{c|}{---}  & \multicolumn{1}{c|}{1.75}  & \multicolumn{1}{cV{4}}{1.71}  \\ \hline
\multicolumn{1}{V{4}c}{40 fs} & \multicolumn{1}{V{4}c}{2.10} & \multicolumn{1}{V{4}c|}{---}  & \multicolumn{1}{c|}{2.02}  & \multicolumn{1}{cV{4}}{1.98}  & \multicolumn{1}{c|}{---}  & \multicolumn{1}{c|}{---}  & \multicolumn{1}{cV{4}}{1.86}  & \multicolumn{1}{c|}{---}  & \multicolumn{1}{c|}{---}  & \multicolumn{1}{cV{4}}{1.63}  \\ \hlineB{4}

\end{tabular}
\end{table}

\begin{table}[htb]
\centering
\caption{Lateral diffusion coefficients $D_\mathrm{lat}$ [10$^{-2}$ nm$^2$/ns] from simulations of DPPC bilayers, obtained on the unwrapped trajectory using \texttt{gmx msd}. The error is smaller than 0.1$\times$10$^{-2}$ nm$^2$/ns. Entries are missing when failed to run properly.}
\label{tab:si-Dlat}
\begin{tabular}{V{4}cV{4}cV{4}cV{4}cV{4}cV{4}cV{4}}
\hlineB{4}
      &    M2   &   M3(-TQ5)   &    M3-SQ5    &   M3-RQ5   &    M3-2$\times$m    \\ \hlineB{4}
20 fs & 5.67    &   5.23       &    5.72      &   5.48     &    5.41             \\ \hline
30 fs & 5.82    &   ---        &    5.47      &   5.69     &    5.07             \\ \hline
40 fs & 5.50    &   ---        &    ---       &   5.41     &    ---              \\ \hlineB{4}
\end{tabular}
\end{table}

\newpage
\section{Diffusion coefficients and quality factors from the GLS estimator}

\begin{figure}[htb]
\begin{center}
    \includegraphics[width=14cm]{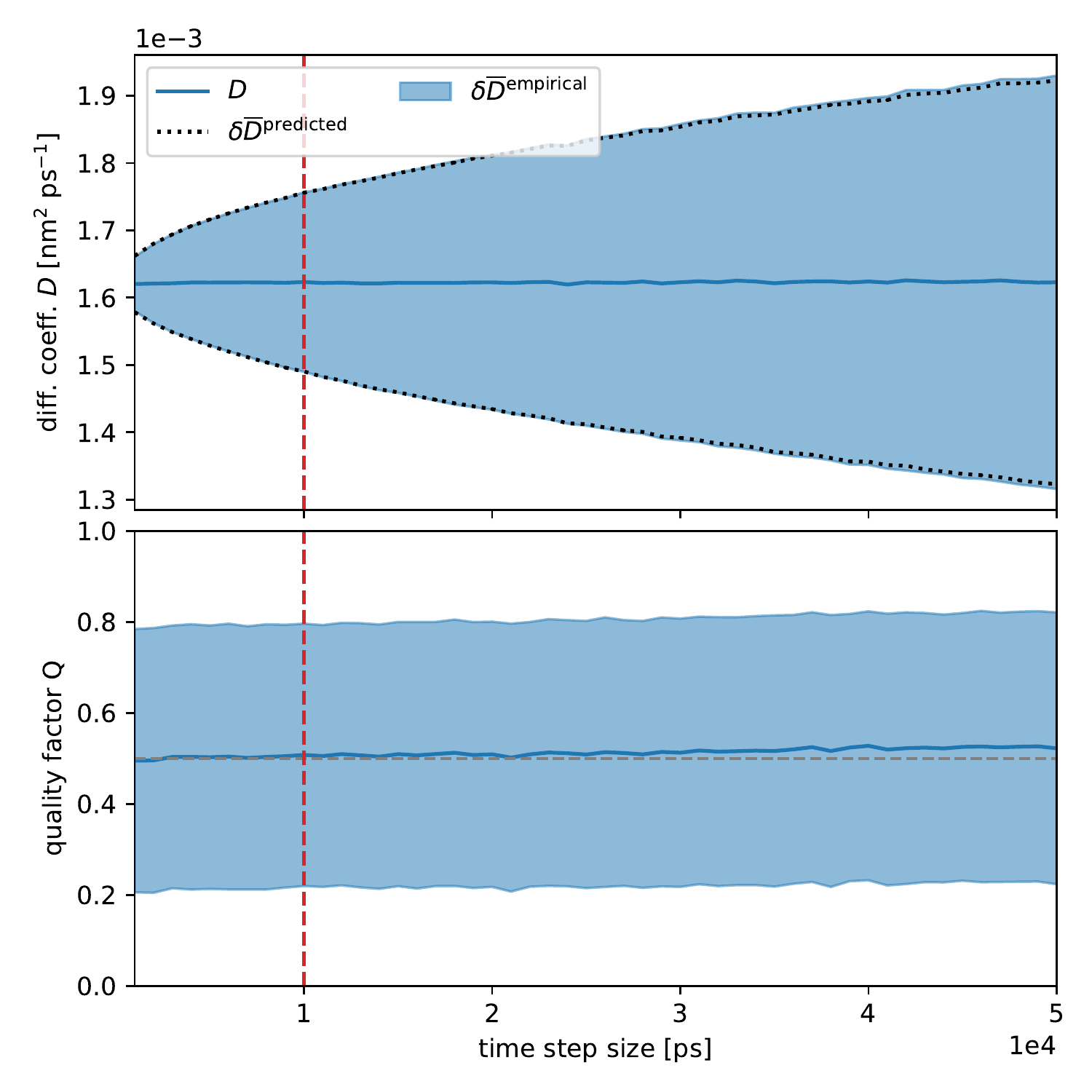}
    \caption{A representative plot of the Generalized Least Squares (GLS) estimation of the water 3D diffusion coefficient of the Martini 3 NaCl solutions. Optimal diffusion coefficients as a function of lag time (TOP). Quality factor a function of lag time (BOTTOM).
    }
    \label{fig:si-solvent-gls}
\end{center}
\end{figure}

\begin{figure}[htb]
\begin{center}
    \includegraphics[width=14cm]{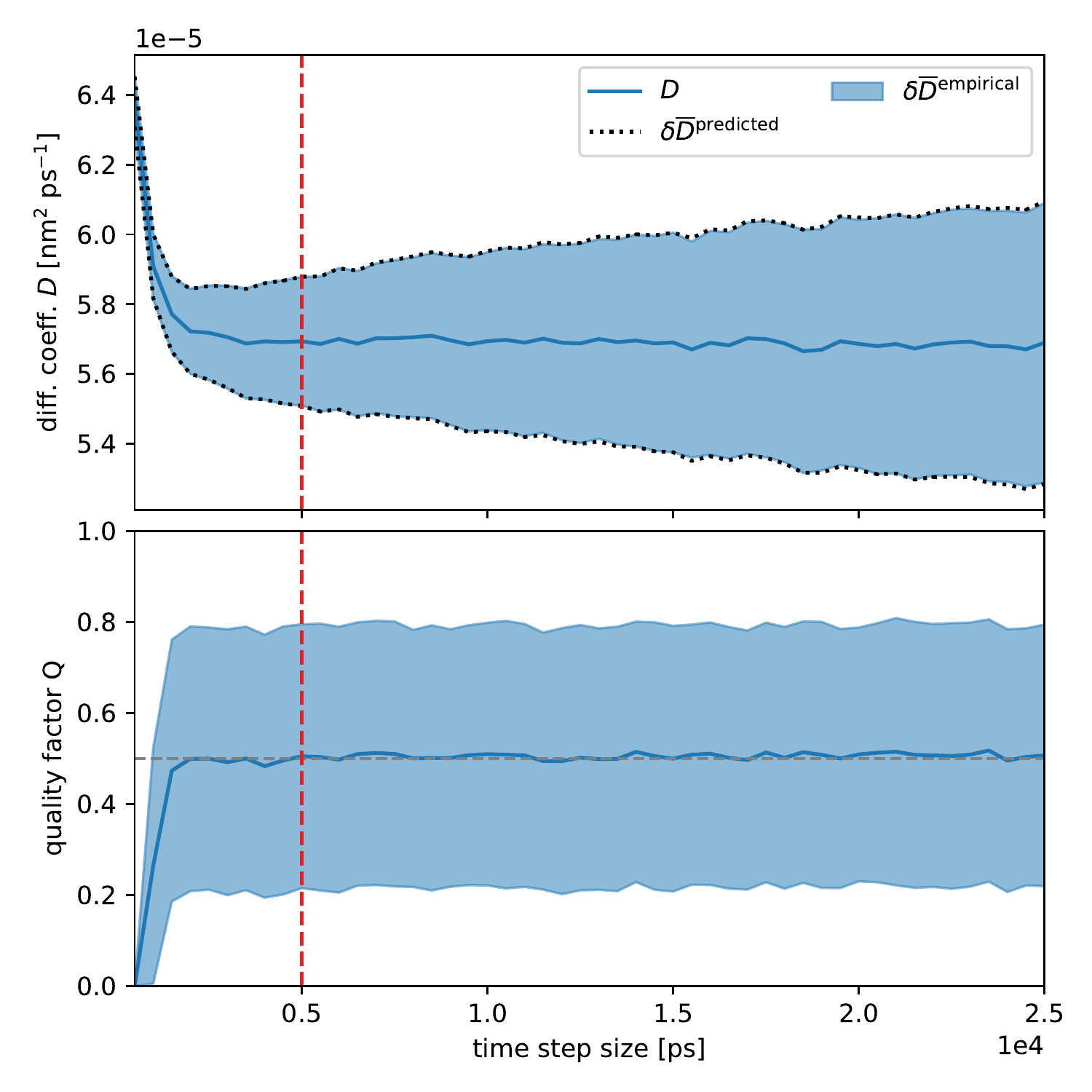}
    \caption{Generalized Least Squares (GLS) estimation of the lateral diffusion coefficient of the Martini 3(-TQ5) DPPC lipids. Optimal diffusion coefficients as a function of lag time (TOP). Quality factor as a function of lag time (BOTTOM).
    }
    \label{fig:si-m3-gls}
\end{center}
\end{figure}

\clearpage
\section{Comparison of diffusion coefficients obtained using \texttt{gmx msd} and the GLS estimator}

\begin{figure}[htb]
\begin{center}
    \includegraphics[width=8.5cm]{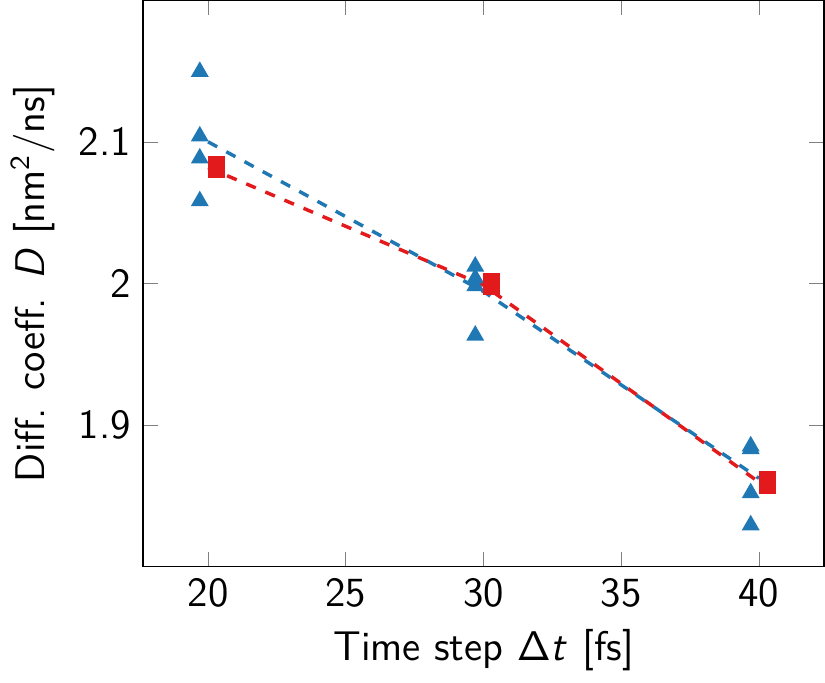}
    \caption{A representative plot of the 3D diffusion coefficients of CG water beads obtained with \texttt{gmx msd} (blue triangles) and the Generalized Least Squares (GLS) estimator (red squares) computed over 4 replicas. The small shifts of the symbols in $\Delta t$ from 20, 30, and 40~fs are for better visibility. The dashed line connect the mean values of the respective data sets, and are guides for the eye.
    }
    \label{fig:si-dcoeff-estimator}
\end{center}
\end{figure}

\newpage
\section{Radial distribution profile of ions in the NaCl solvent simulations}

\begin{figure}[htb]
\begin{center}
    \includegraphics[width=8.5cm]{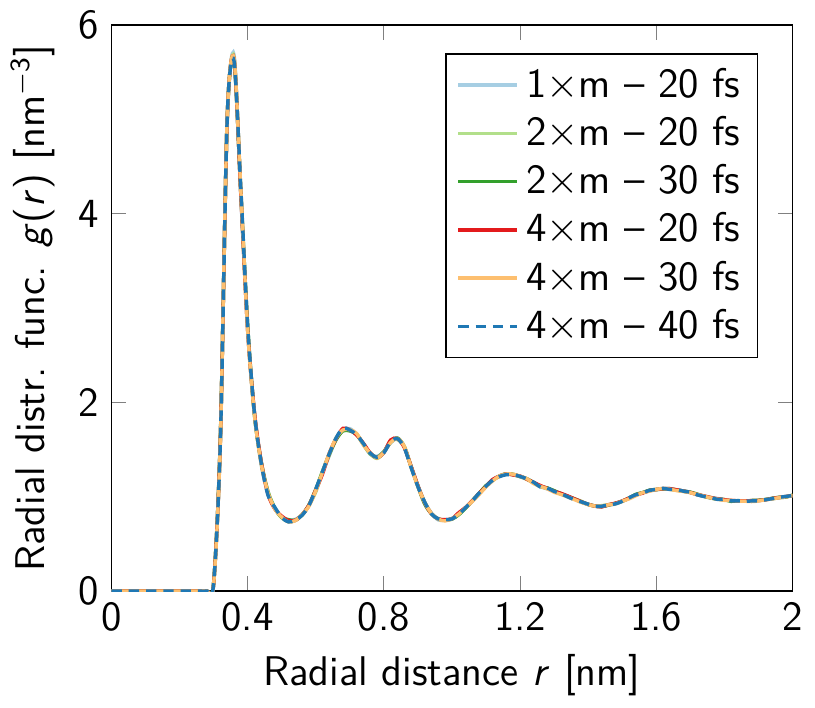}
    \caption{Radial distribution function of ions in the 0.3~M NaCl system, representative of all solvent systems. These functions are identical within sampling error, as equilibrium properties are unaffected by changes in the ionic masses.
    }
    \label{fig:si-bulk-rdf}
\end{center}
\end{figure}

\newpage
\section{Individual density profiles of Na$^+$ and Cl$^-$ ions around the DPPC bilayer}

\begin{figure}[htb]
\begin{center}
    \includegraphics[width=16cm]{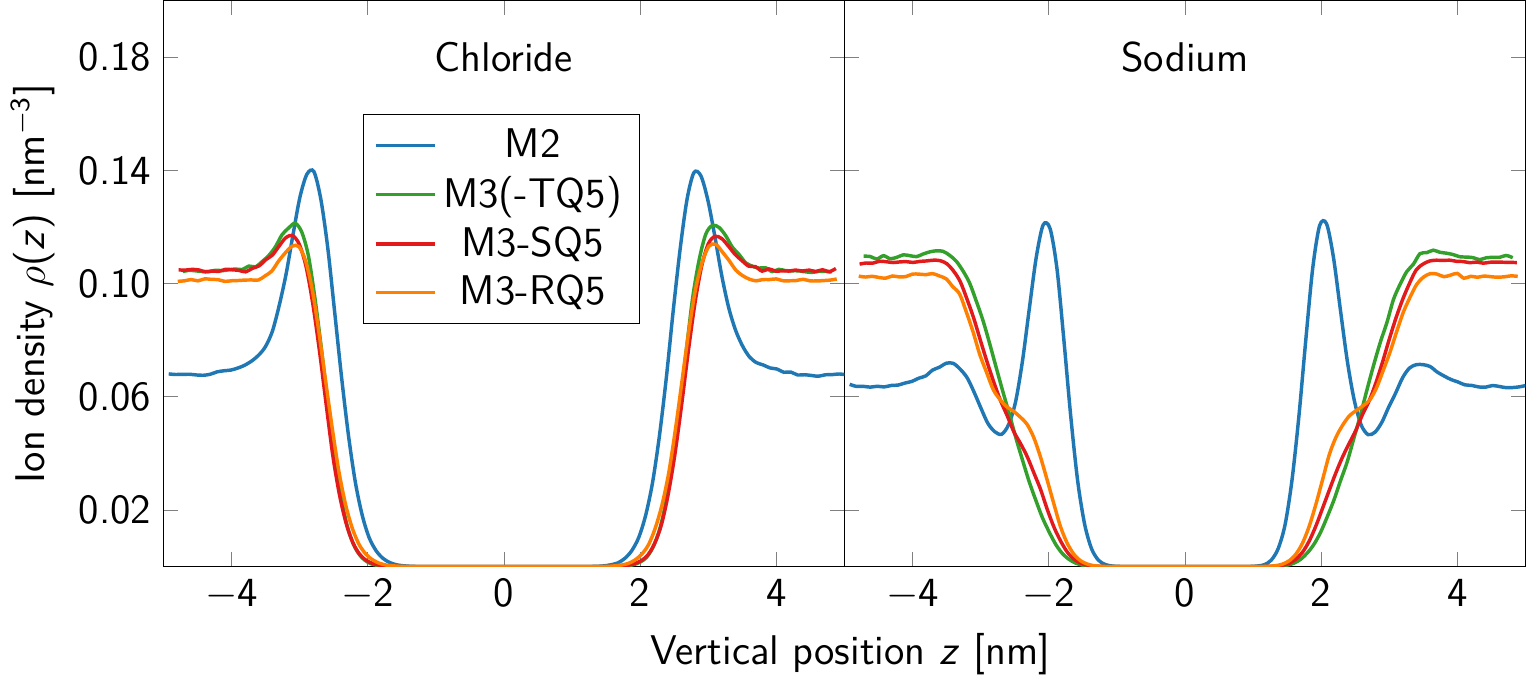}
    \caption{Number density profile of Na$^+$ (LEFT) and Cl$^-$ (RIGHT) ions along the direction normal to the neat DPPC membrane. The mid-plane of the bilayer is located at $z = 0$~nm. Changing the size of the ions has a mild effect on the distributions relative to switching from Martini 2 to Martini 3. The density profiles are equilibrium structural quantities, independent of simulation time step and ionic mass.
    }
    \label{fig:si-membrane-densprof-separate}
\end{center}
\end{figure}

\newpage
\section{PIP\textsubscript{2} containing membranes}

The protocol for simulation and analysis of DPPC bilayers containing 5\% PIP\textsubscript{2} (phosphatidylinositol 4,5-bisphosphate) was identical to the neat DPPC membranes. However, none of the simulations ran at 40~fs time steps, most likely due to the parametrization of PIP\textsubscript{2}. At 30~fs, all systems experienced some failures, but in the case of M3-SQ5, M3-RQ5 and M3-2$\times$m, no more than 2 restarts were needed to obtain a complete trajectory consisting of 500 million steps. On the other hand, M3(-TQ5) suffered from overly frequent crashes similar to the neat DPPC system.

\begin{table}[htb]
\centering
\caption{Average area-per-lipid $A_l$ [\angstrom$^2$] from simulations of PIP\textsubscript{2}-containing DPPC bilayers. The analysis only included the PO4 bead of DPPC, hence all values are higher than in the neat systems. The error is less than $\pm 0.5$~$\angstrom^2$. Entries are missing where simulations failed to run properly.}
\label{tab:si-pip2-apl}
\begin{tabular}{V{4}cV{4}cV{4}cV{4}cV{4}cV{4}}
\hlineB{4}
      & M3(-TQ5)   & M3-SQ5  & M3-RQ5  &    M3-2$\times$m \\ \hlineB{4}
20 fs &  63.1      &  63.0    &  63.0    &    63.1          \\ \hline
30 fs &   ---      &  62.8    &  62.9    &    62.9          \\ \hlineB{4}
\end{tabular}
\end{table}

\begin{table}[htb]
\centering
\caption{Membrane thickness $d$ [nm] from simulations of PIP\textsubscript{2}-containing DPPC bilayers. The analysis only included the PO4 bead of DPPC. The error is less than 0.025~nm. Entries are missing where simulations failed to run properly.}
\label{tab:si-pip2-thick}
\begin{tabular}{V{4}cV{4}cV{4}cV{4}cV{4}cV{4}}
\hlineB{4}
      &   M3(-TQ5)     &    M3-SQ5    &   M3-RQ5     &    M3-2$\times$m    \\ \hlineB{4}
20 fs &   4.161        &     4.164    &   4.168      &    4.161            \\ \hline
30 fs &   ---          &     4.172    &   4.175      &    4.168            \\ \hlineB{4}
\end{tabular}
\end{table}

\begin{table}[htb]
\centering
\caption{Area compressibility modulus $K_\mathrm{A}$ [mN/m] from simulations of PIP\textsubscript{2}-containing DPPC bilayers. The error is less than 2~mN/m. Entries are missing where simulations failed to run properly.}
\label{tab:si-pip2-compress}
\begin{tabular}{V{4}cV{4}cV{4}cV{4}cV{4}cV{4}}
\hlineB{4}
      &   M3(-TQ5)   &    M3-SQ5    &   M3-RQ5   &    M3-2$\times$m    \\ \hlineB{4}
20 fs &   217        &     216      &   216      &    217              \\ \hline
30 fs &   ---        &     200      &   199      &    200              \\ \hlineB{4}
\end{tabular}
\end{table}

\begin{table}[htb]
\centering
\caption{Average order parameters $S_n$ of the DPPC carbon chains from simulations of PIP\textsubscript{2}-containing DPPC bilayers. The error is less than 0.001. Entries are missing where simulations failed to run properly.}
\label{tab:si-pip2-order}
\begin{tabular}{V{4}cV{4}cV{4}cV{4}cV{4}cV{4}}
\hlineB{4}
      &   M3(-TQ5)   &    M3-SQ5      &   M3-RQ5     &    M3-2$\times$m    \\ \hlineB{4}
20 fs &   0.474      &     0.475      &   0.475      &    0.473            \\ \hline
30 fs &   ---        &     0.479      &   0.479      &    0.477            \\ \hlineB{4}
\end{tabular}
\end{table}

\begin{table}[]
\centering
\caption{Lateral diffusion coefficients $D_\mathrm{lat}$ [10$^{-2}$ nm$^2$/ns] of DPPC lipids from simulations of DPPC bilayers containing 5\% PIP\textsubscript{2}. The error is smaller than 0.1$\times$10$^{-2}$ nm$^2$/ns. Entries are missing where simulations failed to run properly.}
\label{tab:si-pip2-dlat}
\begin{tabular}{V{4}cV{4}cV{4}cV{4}cV{4}cV{4}}
\hlineB{4}
      &   M3(-TQ5)   &    M3-SQ5    &   M3-RQ5    &    M3-2$\times$m     \\ \hlineB{4}
20 fs &   5.31       &    5.36      &   5.42      &    5.27              \\ \hline
30 fs &   ---        &    5.23      &   5.26      &    5.14              \\ \hlineB{4}
\end{tabular}
\end{table}

\begin{figure}[htb]
\begin{center}
    \includegraphics[width=8.7cm]{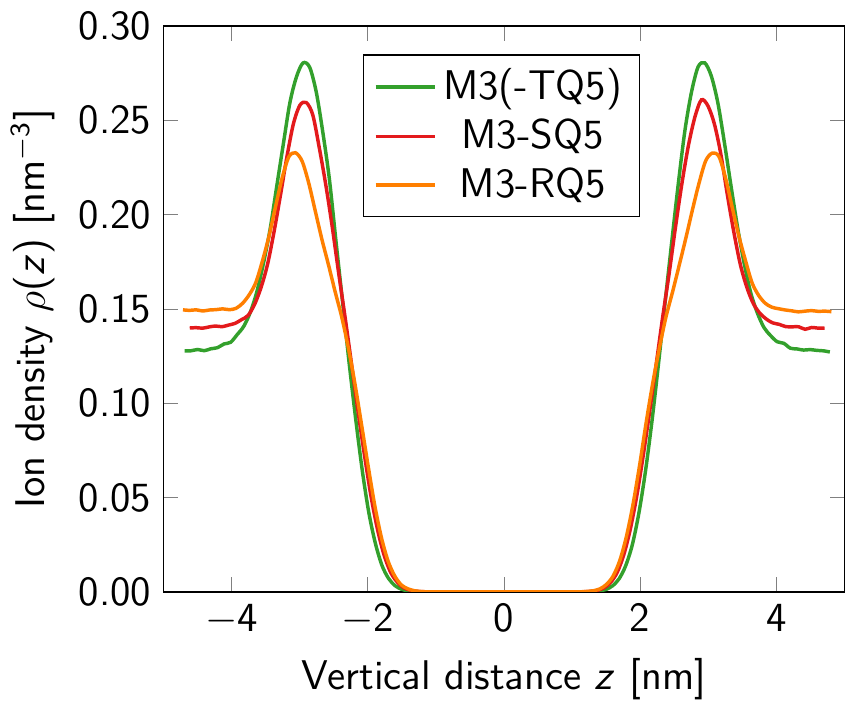}
    \caption{Number density profile of ions along the direction normal to the membrane in PIP\textsubscript{2} containing DPPC bilayers. The mid-plane of the bilayer is located at $z = 0$~nm. As equilibrium properties are not affected by either changes in the ionic mass or time step, the curves correspond only to the 20~fs time step systems and the standard ionic masses.
    }
    \label{fig:si-pip2-densprof}
\end{center}
\end{figure}

\begin{figure}[htb]
\begin{center}
    \includegraphics[width=8.7cm]{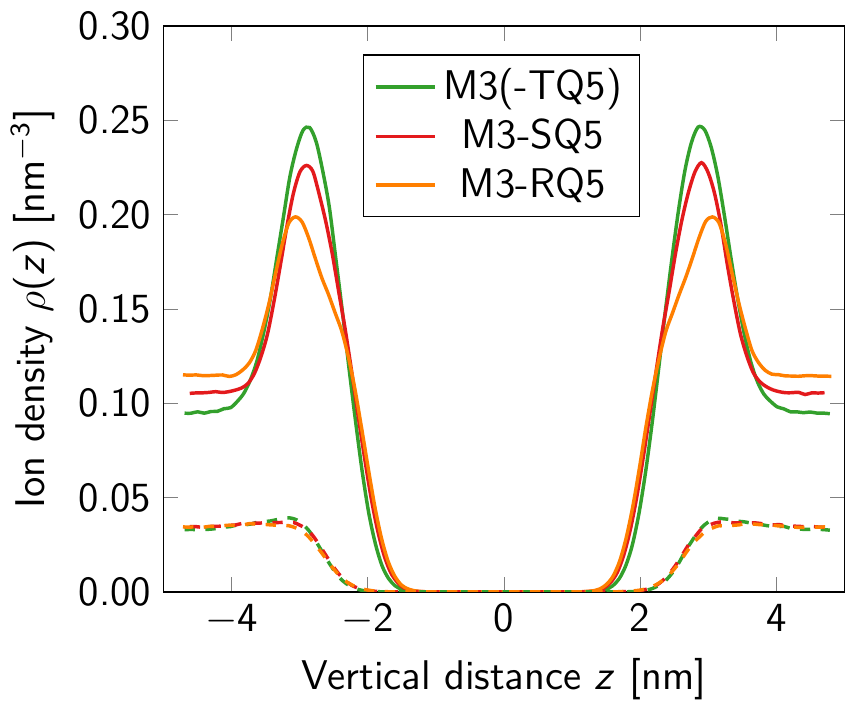}
    \caption{Number density profile of Na$^+$ (solid) and Cl$^-$ (dashed) ions along the direction normal to the PIP\textsubscript{2}-containing DPPC membrane. The mid-plane of the bilayer is located at $z = 0$~nm. As equilibrium properties are not affected by either changes in the ionic mass or time step, the curves correspond only to the 20~fs time step systems and the standard ionic masses.
    }
    \label{fig:si-pip2-densprof-separate}
\end{center}
\end{figure}

\clearpage

\makeatletter\@input{a.tex}\makeatother
\makeatletter\@input{b.tex}\makeatother